\documentclass[twocolumn,preprintnumbers,amsmath,amssymb]{revtex4}

\usepackage{graphicx}% Include figure files
\usepackage{dcolumn}% Align table columns on decimal point
\usepackage{bm}% bold math
\usepackage{amsmath,mathrsfs}

\begin{document}
\title{Angle dependence of the orbital magnetoresistance in bismuth}
\author{Aur\'elie Collaudin$^{1}$, Beno\^{\i}t Fauqu\'e$^{1}$, Yuki Fuseya$^{2}$ Woun Kang$^{3}$ and Kamran Behnia$^{1}$}
\affiliation{(1) Laboratoire de Physique Et d'Etude des Mat\'eriaux (UPMC-CNRS-ESPCI) 10 Rue Vauquelin, 75005 Paris, France\\
(2) Department of Engineering Science, University of Electro-Communications, Chofu, Tokyo 182-8585, Japan\\
$^{3}$ Department of Physics, Ewha Womans University, Seoul 120-750, Korea}

\date {March 18, 2015}

\begin{abstract}
We present an extensive study of angle-dependent transverse magnetoresistance in bismuth, with a magnetic field  perpendicular to the applied electric current and rotating in three distinct crystallographic planes. The observed angular oscillations are confronted with the expectations of semi-classic transport theory for a multi-valley system with anisotropic mobility and the agreement allows us to quantify the components of the mobility tensor for both electrons and holes. A quadratic temperature dependence is resolved. As Hartman argued long ago,  this indicates that inelastic resistivity in bismuth is dominated by carrier-carrier scattering. At low temperature and high magnetic field, the threefold symmetry of the lattice is suddenly lost. Specifically, a $2\pi/3$ rotation of magnetic field around the trigonal axis modifies the amplitude of the magneto-resistance below a field-dependent temperature. By following the evolution of this anomaly as a function of temperature and magnetic field, we mapped the boundary in the (field, temperature) plane separating two electronic states. In the less-symmetric state, confined to low temperature and high magnetic field, the three Dirac valleys cease to be rotationally invariant. We discuss the possible origins of this spontaneous valley polarization, including a valley-nematic scenario.

\end{abstract}


\maketitle

\section{Introduction}
Electric conduction in solids is affected by the application of magnetic field in a variety of ways. The most prominent is orbital magnetoresistance, which is the enhancement of resistivity due to the Lorentz force suffered by charged carriers in presence of magnetic field. As early as 1928, Kapitza discovered that the electric resistivity of bismuth increases by many orders of magnitude upon the application of a large magnetic field\cite{kapitza1928}. The large orbital magnetoresistance is one manifestation of the extreme mobility of carriers in bismuth, itself a consequence of the lightness of electrons, the ultimate reason behind the singular role played by this elemental semi-metal in the history of scientific exploration of electrons in metals\cite{edelman1976}.

During the last few years, bismuth has attracted new attention (A recent review can be found in ref.\cite{fuseya2014}).  A number of intriguing observations on bismuth crystals exposed to strong magnetic fields have have been reported\cite{behnia2007,luli2008,fauque2009,fauque2009b,yang2010}. The angle-resolved Landau spectrum has been found to become exceptionally complex in high magnetic field. In spite of this complexity (and in contrast to what was initially thought\cite{behnia2007,luli2008}), the spectrum resolved by experiment\cite{luli2008,yang2010,zhu2011,zhu2012a,kuechler2014} is in agreement with theoretical expectations\cite{sharlai2009,alicea2009,zhu2011,zhu2012a} based on the band structure of the system and its fine details.

Open questions remain however. The three small pockets of Fermi surface residing at the L-point of the Brillouin zone host Dirac fermions with an extremely anisotropic mass becoming as small as one-thousandth of the free electron mass along the bisectrix axis. These three Dirac valleys are interchangeable upon a $2\pi/3$ rotation around the trigonal axis. But, according to two sets of experimental studies\cite{zhu2012b,kuechler2014}, the three Dirac valleys become inequivalent at low temperature and high magnetic field. The origin of this spontaneous loss of threefold symmetry is yet to be understood.

Metals hosting a small concentration of high-mobility carriers and displaying a large magnetoresistance have attracted much recent attention. In dilute metals such as WTe$_{2}$\cite{ali2014} or Cd$_3$As$_2$\cite{liang2014}, resistivity enhances  by many orders of magnitude upon the application of a magnetic field of 10 T. This is also the case of bismuth\cite{fauque2009,fauque2009b,kopelevich2003} and graphite\cite{kopelevich2003,du2005}, two well-known semi-metals. Both the amplitude of magnetoresistance and its field dependence have been put under scrutiny and are explored and discussed by experimentalists and theorists.

In this paper, we present an extensive study of transverse angle-dependent magnetoresistance in bismuth and establish a detailed map of transverse magnetoresistance for all possible orientations of magnetic field from room temperature down to 2K and up to a magnetic field of 14 T. The study aims to address two distinct issues. The first concerns  the amplitude of magnetoresistance in a compensated semi-metal such as bismuth. Our results show that in any real material, the knowledge of all components of the mobility tensor is required to compute the magnitude of magnetoresistance. The emergence of a valley-polarized state  is the second issue addressed in this study.

Numerous studies of angular oscillations of magnetoresistance in strongly-correlated electron systems have been reported in the past. For three case studies, see Ref.\cite{bergemann2000,hussey2003,kang2007}. These investigations used angular magnetoresistance as a probe, knowing that it shows an extremum when the peculiar topology of the Fermi surface modifies the dimensionality of the cyclotron orbit at a 'magic' angle (dubbed  either Lebed or Yamaji, after those who conceived these commensurability effects). Our experimental configuration is different.  The magnetic field is rotated in a way to keep the magnetic field, $ \vec{B}$, and charge current, $ \vec{j}$,  perpendicular to each other and  the transverse magnetoresistance is measured. In this configuration, the macroscopic Lorentz force between $ \vec{B}$ and $ \vec{j}$, is constant. If the mobility were a scalar, no angular variation would arise. Large angular oscillations are visible even at room temperature and in fields as small as 0.7T in bismuth\cite{zhu2012b}, because carriers have drastically anisotropic mobilities.

Our results provide an opportunity to test the semi-classic transport theory in a particularly constraining format. Since the structure of the three electron ellipsoids and the single hole ellipsoid in bismuth are well-known, one can attempt to fit the experimental data by assuming reasonable values for the components of the mobility tensors of electrons and holes and establish their temperature and field-dependence. We find that, save for a number of important details, our experimental results are in reasonable agreement with the expectations of the semi-classic theory. This paves the way to a quantitative understanding of transverse magnetoresistance in bismuth with obvious implications for other semi-metals and dilute metals.

We also present a detailed study of the configuration in which the current is applied along the trigonal axis and the magnetic field is rotating in the (binary, bisectrix) plane.  We confirm the loss of threefold symmetry at low temperature and high magnetic field previously reported by transport\cite{zhu2012b} and thermodynamic studies \cite{kuechler2014} and find a boundary in the (field, temperature) plane separating two electronic states. The underlying threefold symmetry of the zero-field crystal lattice is lost in the low-temperature-high-field state, but is kept in the high-temperature-low-field state. A phase transition between these two states is clearly detectable at low magnetic field. With increasing magnetic field, the transition shifts to higher temperature and becomes broader. We discuss possible origins of this phase transition and consider the available theoretical scenarios invoking  Coulomb interaction among electrons \cite{abanin2010} or lattice distortion induced by magnetic field reminiscent of Jahn-Teller effect\cite{mikitik2014}. None of the currently available pictures provide an adequate description of the whole range of experimental facts.

\section{Experimental}
\begin{figure*}\centering
\resizebox{!}{0.6\textwidth}
{\includegraphics{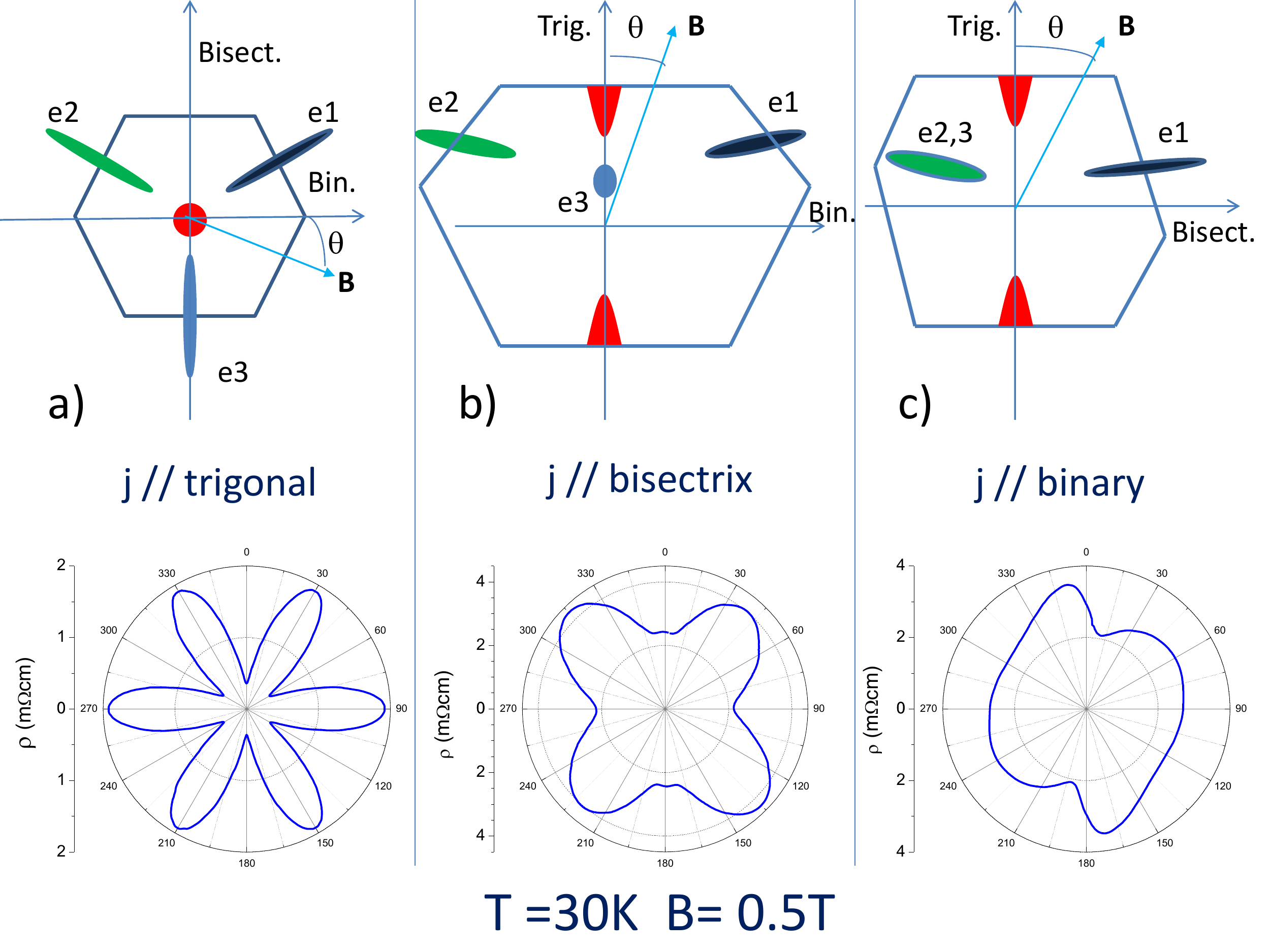}}
\caption{Polar plots of transverse magnetoresistance as a function of the orientation of the magnetic field at T=30K and B=0.5 K. The three plots show the data for three perpendicular planes, Left: field rotates in the (binary, bisectrix) plane; Middle: field rotates in the (binary, trigonal) plane and Right: field rotates in the (biscetrix, trigonal) plane. The electric current is always applied along the crystal axis perpendicular to the rotating plane. Upper panels show the projection of the  Brillouin zone, as well as the hole and electron pockets of the Fermi surface in each rotating plane.}
\end{figure*}

Most measurements were performed using a Quantum Design PPMS apparatus. These studies were complemented with measurements in Seoul using a two-axis home-made set-up in particular to check for any artifact resulting from misalignment. Three bismuth single crystals were cut in a cuboid shape with typical dimensions of (3 $\times$ 4 $\times$ 5 mm$^{3}$) and were used for mapping the magnetoresistance in three perpendicular planes. The typical residual resistivity of these crystals was  1 $\mu\Omega $ cm. Taking a carrier concentration of $n= 3\times10^{17}$cm$^{-3}$ for both hole-like and electron-like carriers and using the simple expression $\sigma_{0}=ne(<\mu_{e}>+<\mu_{h}>)$, the zero-field conductivity would imply that the sum of the average mobility of electrons and holes is as large as $<\mu_{e}>+<\mu_{h}>= 2\times 10^{7}$ cm$^{2}$V$^{-1}$s$^{-1}$. As we will see below, the order of magnitude is confirmed by our magnetoresistance data, but the mobility is very different for electrons and holes and along different orientations. Samples of various cross sections (circular, square, triangular) were studied in order to check the effect of sample geometry on the loss of threefold symmetry.  No significant difference between crystals of different origin (commercial \emph{vs.} home-grown) was observed.

\section{Angular oscillations of transverse magnetoresistance in three perpendicular planes}

The Fermi surface of bismuth consists of a hole ellipsoid and three electron ellipsoids\cite{liu1995}. The hole ellipsoid has a longer axis, which is three times longer than the two other shorter axes and lies along the trigonal axis. The longer axis of each electron ellipsoid  is about 14 times longer than the two shorter axes and lies in a (bisectrix, trigonal) plane, slightly tilted off the bisectrix axis. In the absence of magnetic field, charge conductivity in bismuth is almost isotropic. But, behind this quasi-isotropy  hides an intricate structure of opposite and compensating anisotropies, which reveals itself by the application of a rotating magnetic field.

Fig. 1 presents polar plots of transverse magnetoresistance as the electric current is applied along one crystal axis and the magnetic field is rotated in the plane perpendicular to the applied current. The figure compares the data obtained for three perpendicular planes at a temperature of 30 K and  a magnetic field of 0.5 T.  The three polar plots illustrate how the structure of the Fermi surface leads to very different patterns in each case.
\begin{figure}
\includegraphics[width=8.5cm]{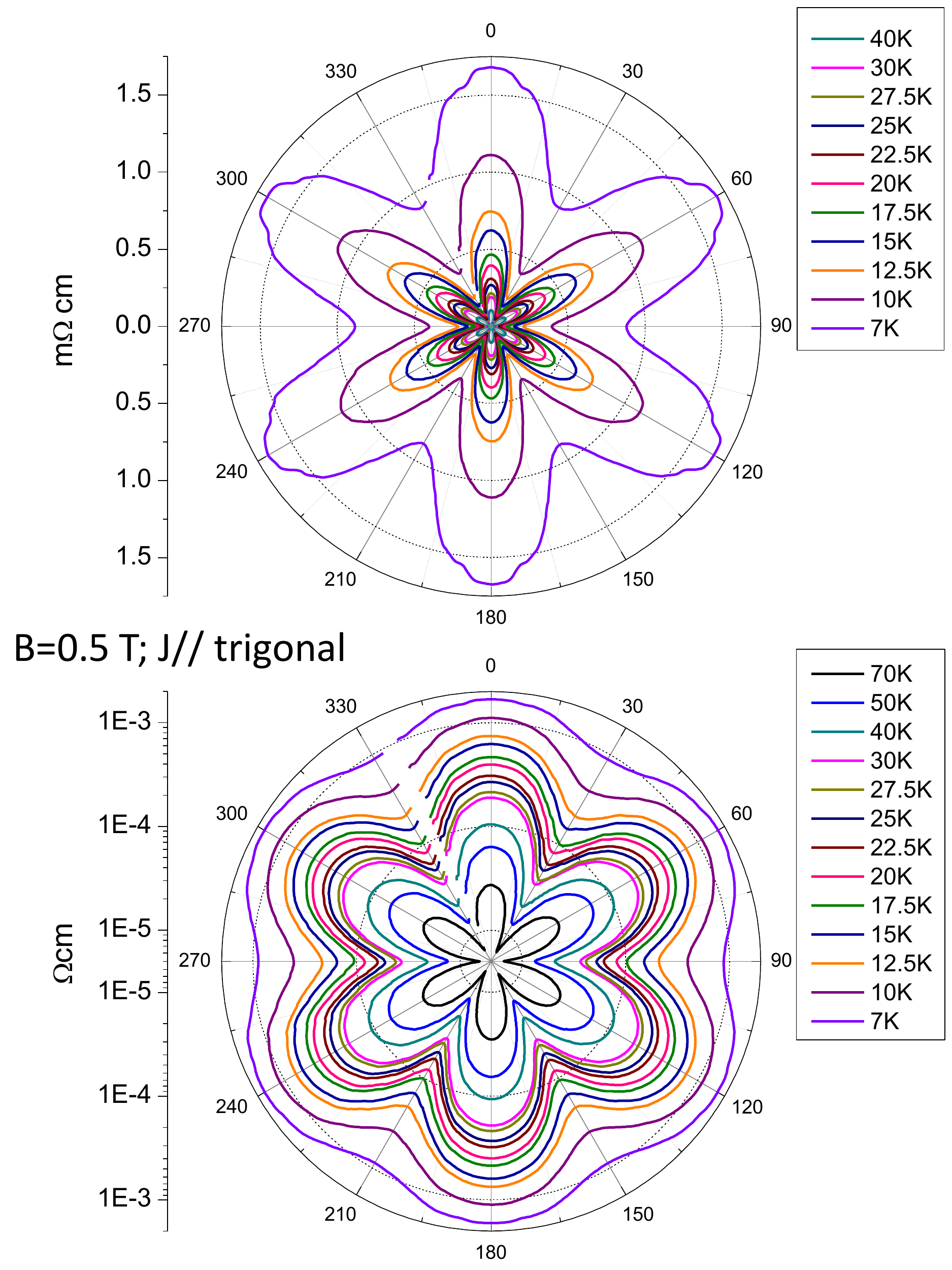}
\caption{Transverse angle-dependent magnetoresistance for a current applied along the trigonal axis and a field rotating in the (binary, bisectrix) plane for different temperatures. The two panels present the same data plotted in  a linear (top) and logarithmic (bottom) scale.}
\end{figure}

When the current is applied along the trigonal axis (Fig. 1a) , magnetoresistance shows six fold angular oscillations. The system has a C$\overline{3}$ symmetry and remains invariant when the magnetic field rotates by 2$\pi$/3. In this configuration, the hole conductivity does not depend on the orientation of magnetic field. On the other hand, electrons show a large variation in their magnetoresistance. It becomes largest when the field is aligned along a bisectrix orientation, since for this orientation of magnetic field, Fermi velocity of electrons  and the microscopic Lorentz force they feel are maximum. When the current is applied along the bisectrix axis and the field rotates in the perpendicular plane (Fig. 1b), magnetoresistance respects two angular symmetries. It is mirror  symmetric (i.e. $\rho (\theta)= \rho(-\theta)$ ) and it keeps the inversion symmetry(i.e. $\rho (\theta)= \rho(\pi +\theta)$ ).  When the current is applied along the binary axis and the field rotates in the third plane(Fig. 1c), only the last [inversion] symmetry is kept. For the last two configuration, one expects a maximal magnetoresistance for holes when the magnetic field is along the trigonal axis. Added to the sum of the contributions of the three electron pockets, this leads to complex patterns. As seen in Fig. 1b and 1c, the total magnetoresistance peaks at intermediate angles off the high-symmetry axes.

\begin{figure}
\includegraphics[width=8.5cm]{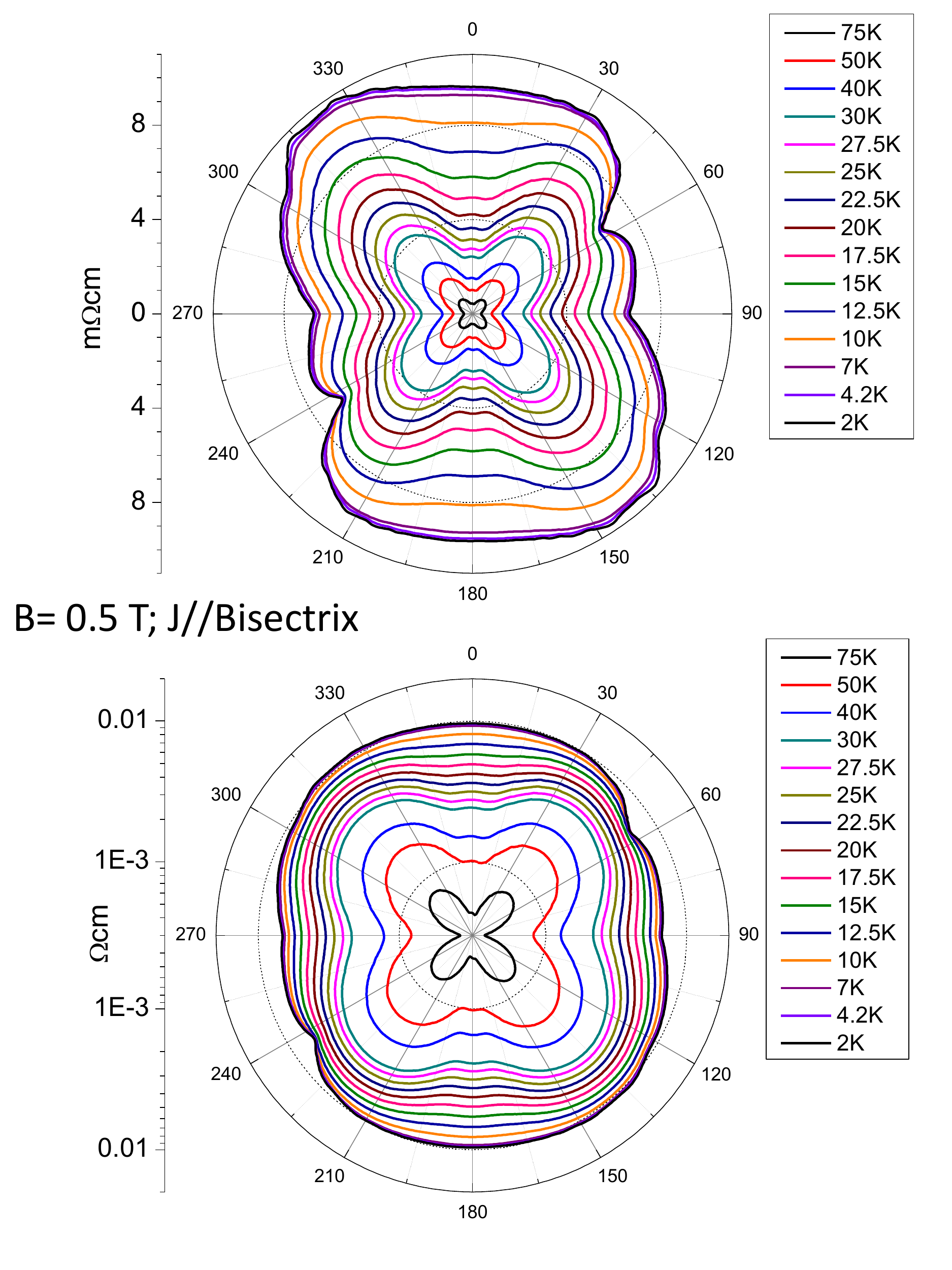}
\caption{Transverse angle-dependent magnetoresistance for a current applied along the bisectrix axis and a field rotating in the (binary, trigonal) plane for different temperatures. The two panels present the same data plotted in  a linear (top) and logarithmic (bottom) scale.}
\end{figure}

These polar plots resemble those reported by Mase, von Molnar and Lawson half a century ago\cite{mase1962}. These authors presented a study of the galvanometric tensor of bismuth for an arbitrary orientation of the magnetic field for a single temperature (20.4 K) and a single magnetic field (0.576 T). The angular variation of our data is in good agreement with their data. However, the absolute amplitude of the magnetoresistance in our data at20 K and 0.5 T, is roughly two times lower.

The extreme sensitivity of magnetoresistance in bismuth upon the orientation of magnetic field permits one to use a rotating magnetic field to tune the contribution of each electron pocket (i.e. each of three valleys) to the total conductivity in the configuration presented in Fig. 1a\cite{zhu2012b}. This provides an interesting opportunity for ``valleytronics''\cite{rycerz2007,behnia2012},  an emerging field of research focused on manipulation of the valley degree of freedom. Such angle-dependent magnetoresistance is expected to be observable in any multi-valley system with anisotropic valleys, as recently demonstrated in the case of SrMnBi$_{2}$\cite{jo2014}. What distinguishes bismuth, however,  is the visibility of such angular oscillations at room temperature and in magnetic fields lower than 1 T\cite{zhu2012b}. This is a consequence of the large mobility of electrons, which exceeds 10$^4$ cm$^{2}$V$^{-1}$s$^{-1}$ at room temperature, and their very anisotropic mass. No solid other than bismuth is currently known to present such properties.

The evolution of the angle-dependent magnetoresistance in bismuth when the field rotates in the (binary, bisectrix) plane was previous reported and the data wa found to be described by an empirical formula for multi-valley conductivity\cite{zhu2012b}. In this paper, extending the measurements to the two other planes, our aim is to see if transverse magnetoresistance in bismuth can be explained for the whole solid angle with a single theoretical model based on a Boltzmann semi-classical picture. We will show that this is indeed the case and the empirical formula used in ref.\cite{zhu2012b} is an approximation of a more general formula for multi-valley systems.

We performed an extended study on several crystals in a wide range of  temperature (2K$<T<$300 K) and magnetic field($B <12$ T). A selection of data for $B = 0.5$ T and $T <75 $ K are presented in Fig. 2-4.  Fig. 2 shows the thermal evolution of the angular magnetoresistance when the current is applied along the trigonal axis and the magnetic field rotates in the (binary, bisectrix) plane. This configuration is identical to the one studied in ref. \cite{zhu2012b} and the results are quite similar. The evolution of the angle-dependent transverse magnetoresistance with decreasing temperature for the two other planes of rotation is shown in Fig. 3 and Fig. 4. In all three cases, the evolution is smooth with increasing magnetoresistance as temperature decreases. This is a consequence of the fact that all components of the mobility tensor enhance with decreasing temperature. At zero magnetic field, the enhancement in mobility with decreasing temperature leads to an enhancement of conductivity. In presence of a field as small as 0.5 T, on the other hand, it leads to an \emph{increase} in resistivity with decreasing temperature. This is because the amplitude of the orbital magnetoresistance is set by the mobility and it exceeds by far the zero-field resistivity of the system in this ``strong-field'' limit.

The thermal evolution of the morphology of the three curves is instructive. For the  first configuration ($I\parallel$ trigonal), there is no visible change in the structure of angle-dependent magnetoresistance with cooling. In the case of the other configurations, a qualitative evolution is observable and new extrema emerge in $\rho(\theta)$ curves as the system is cooled down. Note also the gradual saturation in amplitude of magnetoresistance below 7 K, indicating that mobility has attained its maximum amplitude in this temperature range.

\begin{figure}
\includegraphics[width=8.5cm]{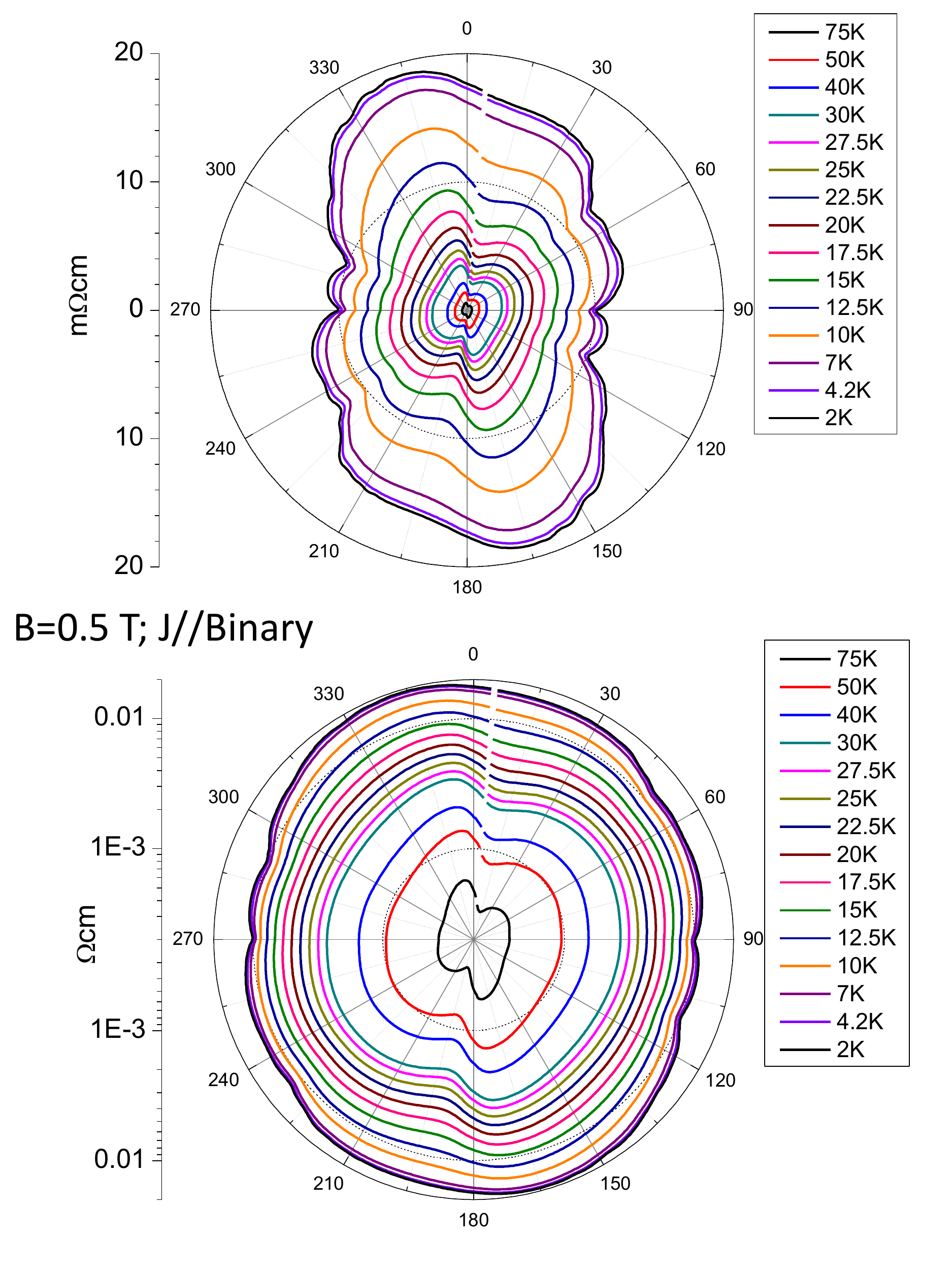}
\caption{Transverse angle-dependent magnetoresistance for a current applied along the binary axis and a field rotating in the (bisectrix, trigonal) plane for different temperatures. The two panels present the same data plotted in  a linear (top) and logarithmic (bottom) scale.}
\end{figure}

For a quantitative treatment of the data, we need to revise the semi-classic transport theory applied to the case of multi-valley system with anisotropic valleys.

\section{The semi-classic transport theory}

Following an earlier work by Abeles and Meiboom\cite{abeles1956}, Aubrey\cite{aubrey1971} wrote the following equation for the charge conduction by a Fermi pocket with a carrier concentration $n$ in presence of electric and magnetic fields:

\begin{equation}
\vec{j} = \hat{\sigma} . \vec{E}=  n e  \hat{\mu} \vec{E} +\hat{\mu} \ \vec{j} \times \vec{B}
\end{equation}

Here, $e$ is the electric charge and  $\hat{\sigma}$  and $\hat{\mu}$ are the conductivity and the mobility tensors. In the absence of magnetic field, Eq. 1 becomes the familiar expression:  $\hat{\sigma}(B=0)= n e \hat{\mu}$.

Abeles and Meiboom\cite{abeles1956} argued that one can derive Eq. 1 starting from the linearized Boltzmann equation, which can be stated in the following manner \cite{Ziman1960}:

\begin{equation}
 e \frac{\partial f^{0}_{k}}{\partial \epsilon_{k}}\overrightarrow{v_{k}}.\vec{E}  + \frac{e}{\hbar}(\overrightarrow{v_{k}}\times \vec{B}) . \overrightarrow{\nabla_{k}g_{k}}=-\frac{g_{k}}{\tau}
\end{equation}

Here, $f^{0}_{k}$ is the Fermi-Dirac distribution in equilibrium, $g_{k}$ represents the deviation from this equilibrium distribution, $\epsilon_{k}$ is the energey and $\overrightarrow{v_{k}}$ the velocity of an electron with a wave-vector $k$ and $\tau$ is the relaxation time. In this picture, the charge current is defined as:

\begin{equation}
\vec{j}= e \int  \overrightarrow{v_{k}}  g_k \overrightarrow{dk}
\end{equation}

Eq. 1 condenses all material-dependent parameters in the mobility tensor defined as:

\begin{equation}
 \hat{\mu} = \tau  \hat{m}^{-1}
\end{equation}

Here  $\hat{m}^{-1}$ is the inverse of the mass tensor. In the simplest case derivation $\tau$ is a scalar. However, this is not necessary and one can easily generalize to a case with different relaxation times along $x$, $y$, $z$.

Aubrey's important contribution was to find that by defining a tensor $ \hat{B}$ whose components are projections of magnetic field along the three perpendicular orientations, one can write a general solution to Eq. 1 in the following manner\cite{aubrey1971}:

\begin{equation}
 \hat{\sigma} = ne \ ( \hat{\mu}^{-1} + \hat{B} )^{-1} \
\end{equation}

The components of the matrix  $ \hat{B}$ are:

\begin{center}
$ \hat{B} = \begin{pmatrix} 0 & -B_3 & B_2 \\ B_3 & 0 & -B_1 \\ -B_2 & B_1 & 0 \end{pmatrix} $
\end{center}

Here, $B_{1}$, $B_{2}$ and $B_{3}$ are the projections of the magnetic field along the three principal axes:
\begin{center}
$ \vec{B} = \begin{pmatrix} B_1 \\ B_2 \\ B_3 \end{pmatrix} $
\end{center}

It is worthy to underline that no assumption has been made on the magnitude of the magnetic field. This is to be contrasted with treatments based on the Jones-Zener expansion, which are only valid in the weak-field limit ($\mu B <1 $)\cite{smith1967}. The Aubrey approach\cite{aubrey1971}, on the other hand, is most appropriate in the strong-field limit ($\mu B >1 $).

Now, to see the physics behind this picture, consider a spherical Fermi surface with an isotropic (i.e. scalar) mobility of $\mu$. In this case, the orientation of the magnetic field has no importance. Let us assume it oriented along the z-axis (that is, $B_{1}=B_{2}$=0 and $B_{3}=B$). In this case, the solution implied by Eq. 5 becomes:

\begin{center}
$ \hat{\sigma} = \begin{pmatrix} \sigma_{\bot} &\sigma_{H} & 0 \\ -\sigma_{H} & \sigma_{\bot} & 0 \\ 0 & 0 & \sigma_{\|} \end{pmatrix} $
\end{center}

 Here, the transverse ($\sigma_{\bot}$), Hall ($\sigma_{H}$) and longitudinal ($\sigma_{\|}$)  components of the conductivity tensor take the following familiar expressions:
\begin{equation}
\sigma_{\bot} = \frac{ne\mu}{1+\mu^{2}B^{2}};
\sigma_{H} = \frac{ne\mu}{1+\mu^{2}B^{2}} \mu B;
\sigma_{\|} = ne\mu
\end{equation}

In other words, magnetic field modifies only the transverse conductivity and leaves the longitudinal conductivity unchanged. Moreover, the Hall component outweighs the transverse magnetoresistance in the high-field ($\mu B >1 $) limit. Now, consider an ellipsoidal Fermi surface, with an anisotropic mobility:
\begin{center}
$ \hat{\mu} = \begin{pmatrix} \mu_1 & 0 & 0 \\ 0 & \mu_2 & 0 \\ 0 & 0 & \mu_3 \end{pmatrix} $
\end{center}

Rotating the magnetic field in the (1, 2) plane and measuring electric conductivity along the third axis, according to Eq. 2, one would expect to find:

\begin{equation}
\sigma_{33} = \frac{ne\mu_{3}}{1+B^{2}\mu_{3}(\mu_{2}\cos^{2}\theta+\mu_{1}\sin^{2}\theta)}
\end{equation}

Here $\theta$ is the angle between the magnetic field and the binary (i.e. 2nd) axis. When $\mu_{1}\gg \mu_{2}$, this equation becomes identical to the empirical formula used to fit the data obtained for the field rotating in the (binary, bisectrix) plane\cite{zhu2012b}.

Let us now turn to a multi-valley system where the total conductivity is the sum of the contributions by each valley. In the specific case of bismuth, one can write:

\begin{equation}
\hat{\sigma}\mathrm{_{tot}} = \underset{i=1-3}{\sum} \hat{\sigma}_i^{e} + \hat{\sigma}^{h} .
\end{equation}

Here $\hat{\sigma}^{h}$ is the conductivity tensor of the hole pocket and $\hat{\sigma}_i^{e}$ is the conductivity tensor of the electron pocket indexed $i$.

The structure of electron ellipsoids in bismuth is complex. They have no circular cross sections and do not lie along any symmetry axis. Their mobility tensor, like their effective mass tensor\cite{dresselhaus1971}, has four distinct and finite components. In the case of the electron ellipsoid which has its longer axis in the ($x$, $z$) plane, this tensor can be expressed as:

\begin{center}
$ \hat{\mu}_1^{e} = \begin{pmatrix} \mu_1 & 0 & 0 \\ 0 & \mu_2 & \mu_4 \\ 0 & \mu_4 & \mu_3 \end{pmatrix} $
\end{center}

The three electron pockets are equivalent to each other through a $2\pi/3$ rotation. Therefore the mobility tensor for the two other electron ellipsoids would be:
\begin{center}
$ \hat{\mu}_2 ^{e}= \hat{R}_{2 \pi /3}^{-1} \ . \ \hat{\mu}_1^{e} \ . \ \hat{R}_{2 \pi /3} $ \; ; \;
$ \hat{\mu}_3^{e} = \hat{R}_{4 \pi /3}^{-1} \ . \ \hat{\mu}_1^{e} \ . \ \hat{R}_{4 \pi /3} $
\end{center}

Here, $ \hat{R}_{\theta} $ is the rotation matrix for a rotation angle of $\theta$ around the trigonal axis :

\begin{center}
$ \hat{R}_{\theta} = \begin{pmatrix} \cos \theta & -\sin \theta & 0 \\ \sin \theta & \cos \theta & 0 \\ 0 & 0 & 1 \end{pmatrix} $
\end{center}

The hole pocket is an ellipsoid with a circular cross section perpendicular to the trigonal axis. It has identical projections in the binary and bisectrix planes. Therefore, there are only two distinct and finite components and its mobility tensor $ \hat{\nu}$ can be written as:

\begin{center}
$ \hat{\nu} = \begin{pmatrix} \nu_1 & 0 & 0 \\ 0 & \nu_1 & 0 \\ 0 & 0 & \nu_3 \end{pmatrix} $ .
\end{center}

Using this formalism, one can compute the components of total conductivity tensor for a given set of seven parameters ($\mu_{1-4}$, $\nu_{1,3}$ and  $n$).  In order to compare with the experiment, one needs to invert the calculated tensor and obtain the relevant component of the resistivity tensor ( $\hat{\rho} = \hat{\sigma}^{-1}$). In our particular case, we have measured resistivity, $\rho_{11}$,  $\rho_{22}$,  $\rho_{33}$,  along three principal axes. The link between $\rho_{ii}$ and the components of the conductivity tensor can be written as:

\begin{equation}
\rho_{ii} = \frac{1}{\sigma_{ii} +\delta\sigma_{ii}}
\end{equation}

Where:
\begin{equation}
\delta\sigma_{11} =\frac{\sigma_{12}\sigma_{23}\sigma_{31}+\sigma_{13}\sigma_{21}\sigma_{32}-\sigma_{22}\sigma_{13}\sigma_{31}-\sigma_{33}\sigma_{12}\sigma_{21}} {\sigma_{22}\sigma_{33}- \sigma_{23}\sigma_{32}}
\end{equation}

\begin{equation}
\delta\sigma_{22} =\frac{\sigma_{12}\sigma_{23}\sigma_{31}+\sigma_{13}\sigma_{21}\sigma_{32}-\sigma_{11}\sigma_{23}\sigma_{32}-\sigma_{33}\sigma_{12}\sigma_{21}} {\sigma_{11} \sigma_{33} - \sigma_{13} \sigma_{31}}
\end{equation}

\begin{equation}
\delta\sigma_{33} =\frac{\sigma_{12}\sigma_{23}\sigma_{31}+\sigma_{13}\sigma_{21}\sigma_{32}-\sigma_{11}\sigma_{23}\sigma_{32}-\sigma_{22}\sigma_{13}\sigma_{31}} {\sigma_{11} \sigma_{22} - \sigma_{12} \sigma_{21}}
\end{equation}
\begin{figure*}\centering
\resizebox{!}{0.6\textwidth}
{\includegraphics{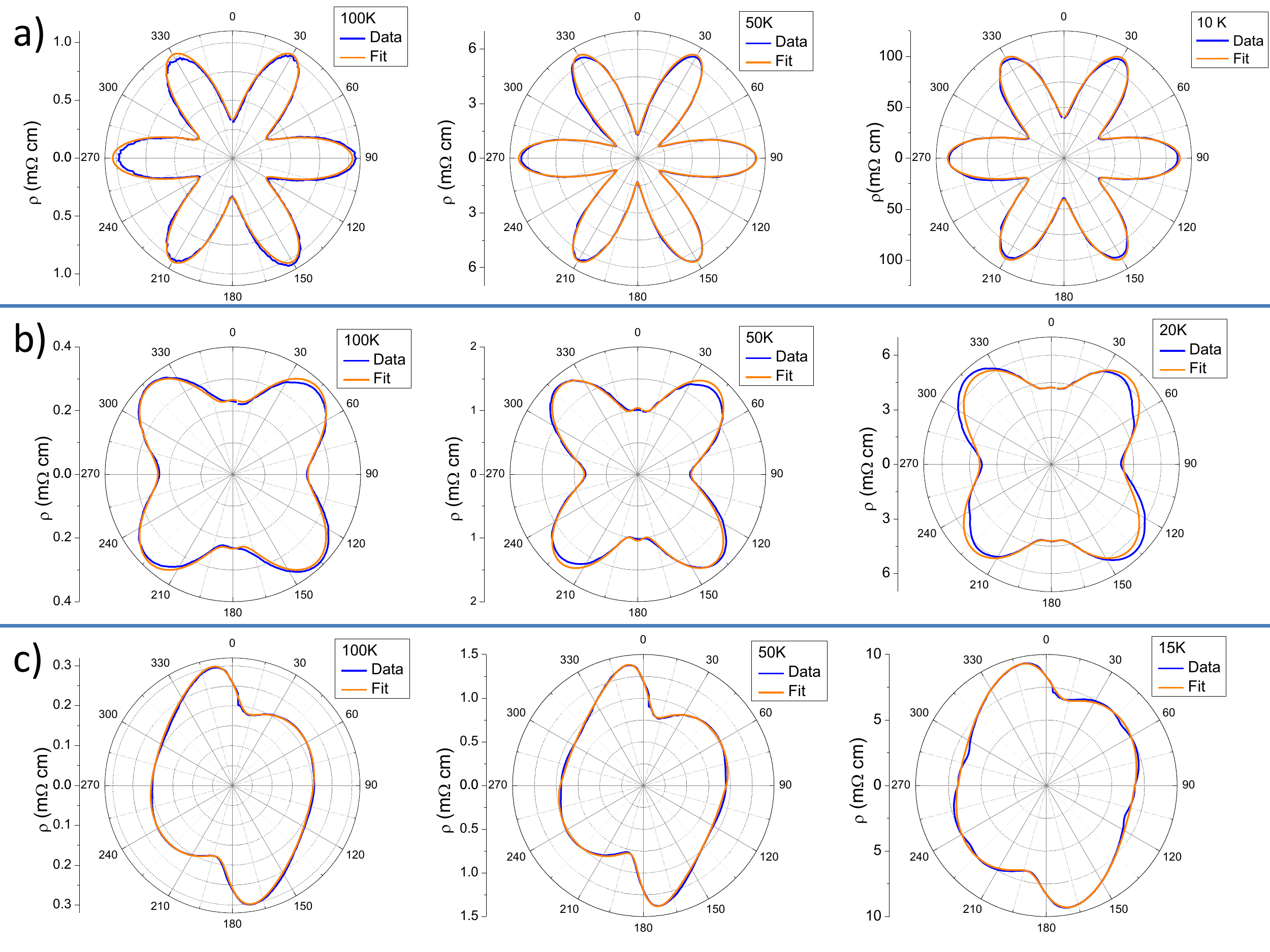}}
\caption{Comparison between the data and the theoretical fit based on the optimal parameters at B=0.5 T and three different temperatures for : a) Magnetic field rotating in the (binary, bisectrix) plane; b) magnetic field rotating in the (trigonal, bisectrix) plane; c) magnetic field rotating in the (trigonal, binary) plane.}
\end{figure*}

One may wonder if there is any room for harmless approximations here. These expressions contain numerous off-diagonal components. Onsager relations imply that $\sigma_{ij}(B)=-\sigma_{ji}(-B) $.  Agkoz and Saunders have detailed the restrictions imposed by Onsager reciprocity relations for all 32 crystallographic point groups\cite{akgoz1975}. They demonstrated that in the case of A7 structure of bismuth and antimony, the off-diagonal components of the conductivity tensor can have both even and odd parts. In particular, when the field is along the binary axis, there is a so-called 'Umkehr' effect, leading to the inequality $\sigma_{23}(B_{1})\neq -\sigma_{32}(B_{1}) $. This means that the first two terms of the nominators in equations 10 to 12 do not cancel out.

It is true that in a compensated semi-metal with an equal concentration of electrons and holes, the opposite Hall response of electrons and holes can cancel each other. In case of negligible off-diagonal components, one would simply find $\rho_{ii} \simeq \sigma_{ii}^{-1}$. When the magnetic field is aligned along each of the three high-symmetry axes, one can experimentally verify that $\rho_{ij}\ll\rho_{ii}$  and thus $\sigma_{ij}\ll\sigma_{ii}$. However, for arbitrary orientations of the magnetic field, the magnitude of the off-diagonal components is large enough to generate non-negligible effects. We found that in order to find a satisfactory fit, the full expressions of equation 9-12 are to be used. In particular, in the case of the third configuration, i.e. the field rotating in the (bisectrix, trigonal) plan], without taking $\delta\sigma_{ii}$ into account, even a qualitative agreement with the experimental data is impossible to obtain.

In 1974, S\"{u}mengen, T\"{u}retken and Saunders\cite{sumengen1974} employed this formalism to quantify the components of the mobility tensor in bismuth using the experimental data published a decade before by Mase and co-workers\cite{mase1962} for a single temperature (20.4 K) and a single magnetic field (0.576 T). They found that experimentally-observed angular variation can be theoretically reproduced and the extracted components of the mobility tensor match those found by measuring components of resistivity, $\rho_{ij}$ in the low-field ($\mu B \ll 1$)limit\cite{hartman1969,michenaud1972}. In the next section, we are going to use this procedure for our extensive set of data.

\section{The mobility tensor}

According to the theoretical frame described in the previous section, the angle-dependent transverse magnetoresistance can be fit using equations 9-12 with seven adjustable parameters. These are the four components of the mobility tensor of electrons, $\mu_{1-4}$, the two components of the mobility tensor of holes, $\nu_{1, 3}$ and the  carrier density of holes, $n_{h}$, which is equal to the sum of the carrier densities of the three electron pockets ($n_{h}=n_{e1}+n_{e2}+n_{e3}$).

We tried such fits and found that good fits with plausible parameters can be achieved. Polar plots comparing the experimental data and the expectations of the theory are shown in  Fig. 5.  This agreement establishes that the large anisotropic magnetoresistance of bismuth can be, mostly if not entirely, explained by the semi-classic theory. At lower temperatures and higher magnetic fields, the fits become less satisfactory.

There are several reasons for the gradual inadequacy of the model as the temperature lowers and the magnetic field increases. First of all, Landau quantization becomes sharp enough to introduce additional angular structure not expected by the model used here. The second reason is the loss of the threefold symmetry, which will be discussed in detail in the following section. A third is the apparent effect of the applied magnetic field on the components of the mobility tensor. This last feature appears only after putting under scrutiny the details of the fitting procedure.

We found that in some cases, two different sets of fitting parameters could both yield satisfactory fits of comparable quality. In order to minimize this uncertainty, we attempted to reduce the number of adjustable parameters by excluding the carrier density and the tilt angle as variable parameters.

It has been known that carrier density in bismuth remains constant up to 50 K and then begins to increase as thermally excited carriers are introduced across the small gap at the L-point\cite{issi1979}. We took a fixed value of $ n_{e}=n_{h}=3\times 10^{17}$ cm$^{-3}$ (the magnitude obtained from de Hass-van Alphen measurements\cite{bhargava1967}) for $T<50$ K and let it evolve as a free parameter for temperatures above.

The tilt angle, $\theta$, between the longer axis of the electron ellipsoid and the crystalline bisectrix axis, generates the fourth mobility component, $\mu_{4}$.  If the relaxation time tensor happens to be aligned along the high-symmetry crystals axes,  the mobility tensor and the mass tensor should have identical tilt angles. It was measured to be 6.4 degrees by de-Haas-van-Alphen effect studies\cite{bhargava1967} and its controversial sign  was settled by refined comparison between the data on structural and electronic properties\cite{brown1968}. The recent study of angle-resolved Landau spectrum\cite{zhu2011} found a tilt angle($\sim 6.2 ^{\circ}$ ) close to the previous reports.

In the case of the mobility tensor, the tilt angle can be  expressed as \cite{hartman1969,sumengen1974}:

\begin{equation}
\theta= \frac{1}{2}\arctan (\frac{2\mu_{4}}{\mu_{2}-\mu_{3}})
\end{equation}

Hartman found that the tilt angle of the mobility tensor is very close to this value and concluded that the relaxation time tensor has little or no tilt\cite{hartman1969}. We found that the best fits to our data point to a value close to $6.8 ^{\circ}$ and in order to reduce the fitting uncertainty imposed this as a constraint to our fitting procedure.

Figure 6 and Figure 7 show  the the temperature-dependence of the mobility components of electrons and holes extracted from our fits to the data at $B=0.5$ T. Our data is compared with zero-field values extracted from the galvanometric coefficients of bismuth in the low-magnetic field limit. Several remarks are in order.

First of all, the sheer magnitude of the mobility of Dirac electrons in bismuth is remarkably large. At 10 K, $\mu_{1}$ in bismuth becomes as large 1000 T$^{-1}$ or 10$^{7}$ cm$^{2}$V$^{-1}$s$^{-1}$. This is slightly lower than what Hartman found at the same temperature. According to his results, $\mu_{1}$ becomes as large as 10$^{8}$ cm$^{2}$V$^{-1}$s$^{-1}$. Therefore, and this may be worth being recalled, electrons in bismuth  are by far more mobile than carriers seen in any other three-dimensional solid\cite{liang2014}.

\begin{figure}
\includegraphics[width=8.5cm]{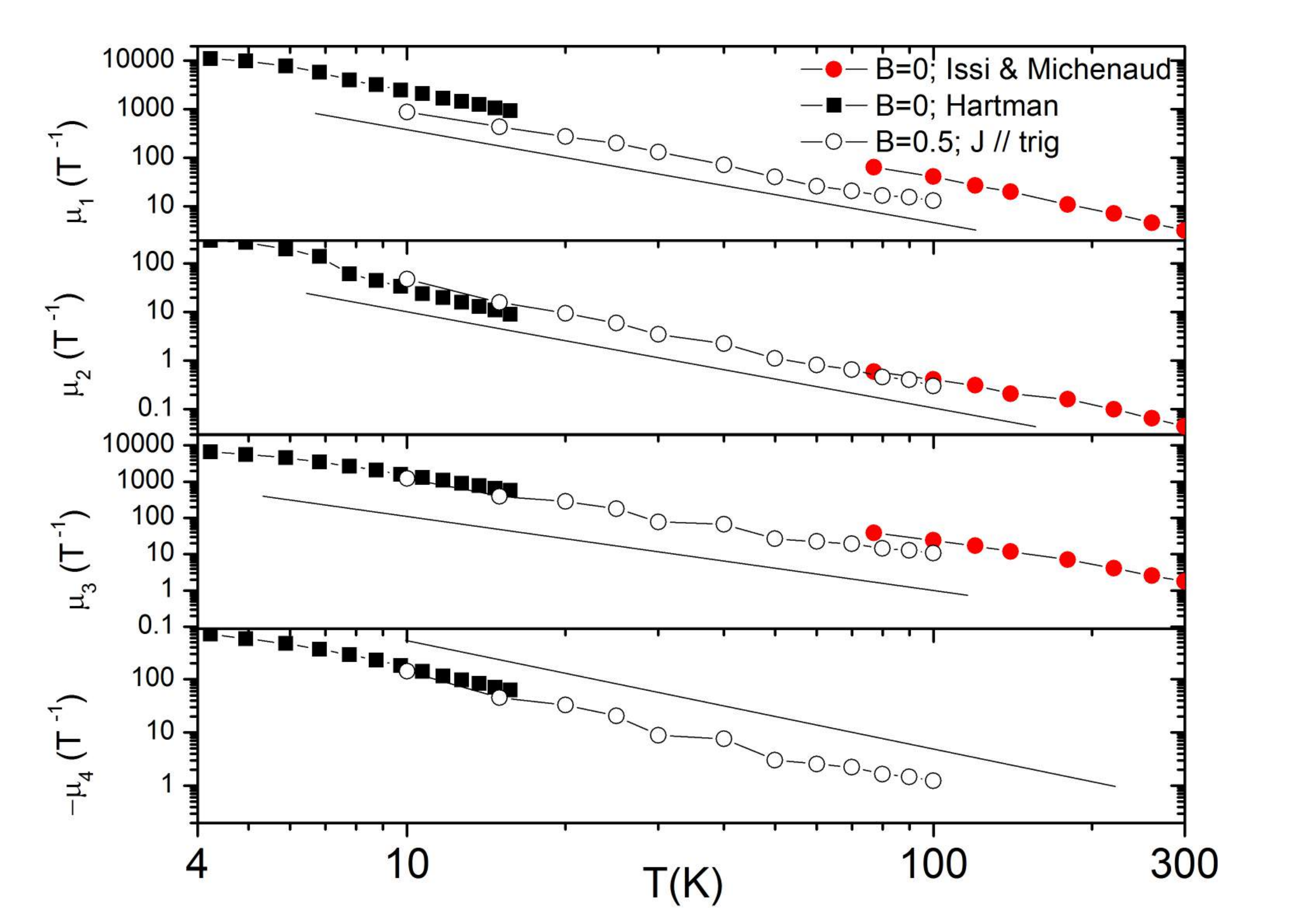}
\caption{Temperature dependence of the components of the electron mobility tensor, $\mu_{i}$, obtained by fitting our angle-dependent magnetoresistance data at $B=0.5 $T for a current applied along the trigonal axis. Also shown are the results reported for zero magnetic field by Hartman below 15 K\cite{hartman1969} and by Michenaud and Issi above 77 K \cite{michenaud1972}. Solid lines represent a T$^{-2}$ temperature dependence (No data on $\mu_{4}$ was reported in the latter case).}
\end{figure}

\begin{figure}
\includegraphics[width=8.5cm]{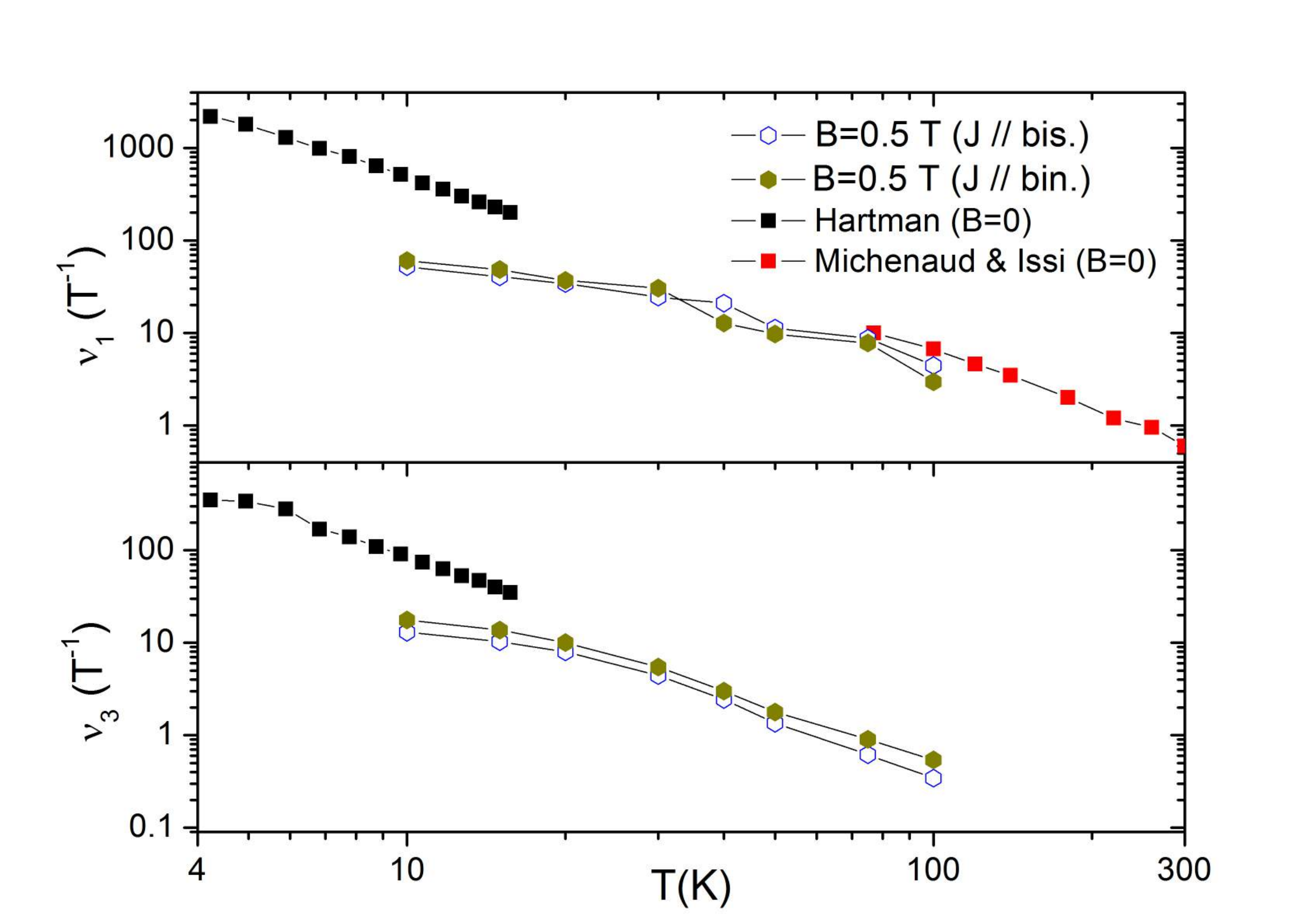}
\caption{Temperature dependence of the components of the hole mobility tensor, $\nu_{i}$,  obtained by fitting our angle-dependent magnetoresistance data at $B=0.5 $T for current applied along the binary or bisectrix axis. Also shown are the results reported for zero magnetic field by Hartman by Hartman below 15 K\cite{hartman1969} and by Michenaud and Issi above 77 K\cite{michenaud1972} (No $\nu_{3}$ was reported in the latter case).}
\end{figure}

Second, the  mobility tensor is anisotropic. Qualitatively, this reflects the anisotropy of the band mass. The anisotropy is larger for electron, which have a more anisotropic mass, than holes. The mobility of electrons is lowest along the bisectrix, which is the direction for which the electrons are heaviest. As for holes, they are lighter and more mobile perpendicular to the trigonal axis than parallel to it. However, the correspondence between the two tensors remains qualitative. The mass anisotropy of electrons, extracted from studies of quantum oscillations\cite{bhargava1967,zhu2011} is as large as 200. The anisotropy of mobility does not exceed 40. This implies an anisotropic relaxation time tempering the mass anisotropy.

It is instructive to compare our mobilities at $B=0.5$ T for various orientations of magnetic field to those reported by previous authors in the zero-field limit\cite{hartman1969,michenaud1972}. As seen in Fig. 6 in the case of electrons, a reasonable match is observed between the different sets of results. In the case of holes, on the other hand, as seen in Fig. 7, a difference is visible at low temperature: the mobilities extracted from  data saturate to values significantly lower than what was found in the zero-field limit. This may be due to a difference in the ultimate carrier mean-free-path in different samples. The residual resistivity in our samples is an order of magnitude larger than those studied by Hartman. It may also partially arise  from a field-induced decrease in the magnitude of $\nu_{1}$ and $\nu_{3}$.

The effect of magnetic field on the various components of mobility tensor shows itself in another fashion. Fig. 8 compares the magnitude of the components of the mobility tensor of electrons for the three rotating planes. As seen in the figure, the  discrepancy remains within the error margin at high temperatures, but exceeds it in the low-temperature regime. Moreover, the discrepancy has a clear pattern. All $\mu_{1-4}$ components become lower in the case of the third configuration, with the current applied along the bisectrix axis and the magnetic field rotating in the (binary, trigonal) plane. This indicates that the components of the mobility tensor are affected by the magnitude and orientation of magnetic field in the low-temperature limit.

Hartman\cite{hartman1969} noticed that the  components of mobility tensor present a quadratic decrease with temperature over a wide temperature window. This T$^{-2}$ temperature dependence of mobility (and consequently zero-field conductivity) is a hallmark of electron-electron scattering.  Long ago, Baber\cite{baber1937} argued that in a metal with multiple electron reservoirs coupled to the lattice thermal bath, electron-electron scattering should give rise to a T$^2$ inelastic resistivity. In the case of metals hosting correlated electrons, this $T^{2}$  resistivity has been extensively documented and is one of the two main ingredients of the much-discussed Kadowaki-Woods ratio\cite{kadowaki1986}.

\begin{figure}
\includegraphics[width=8.5cm]{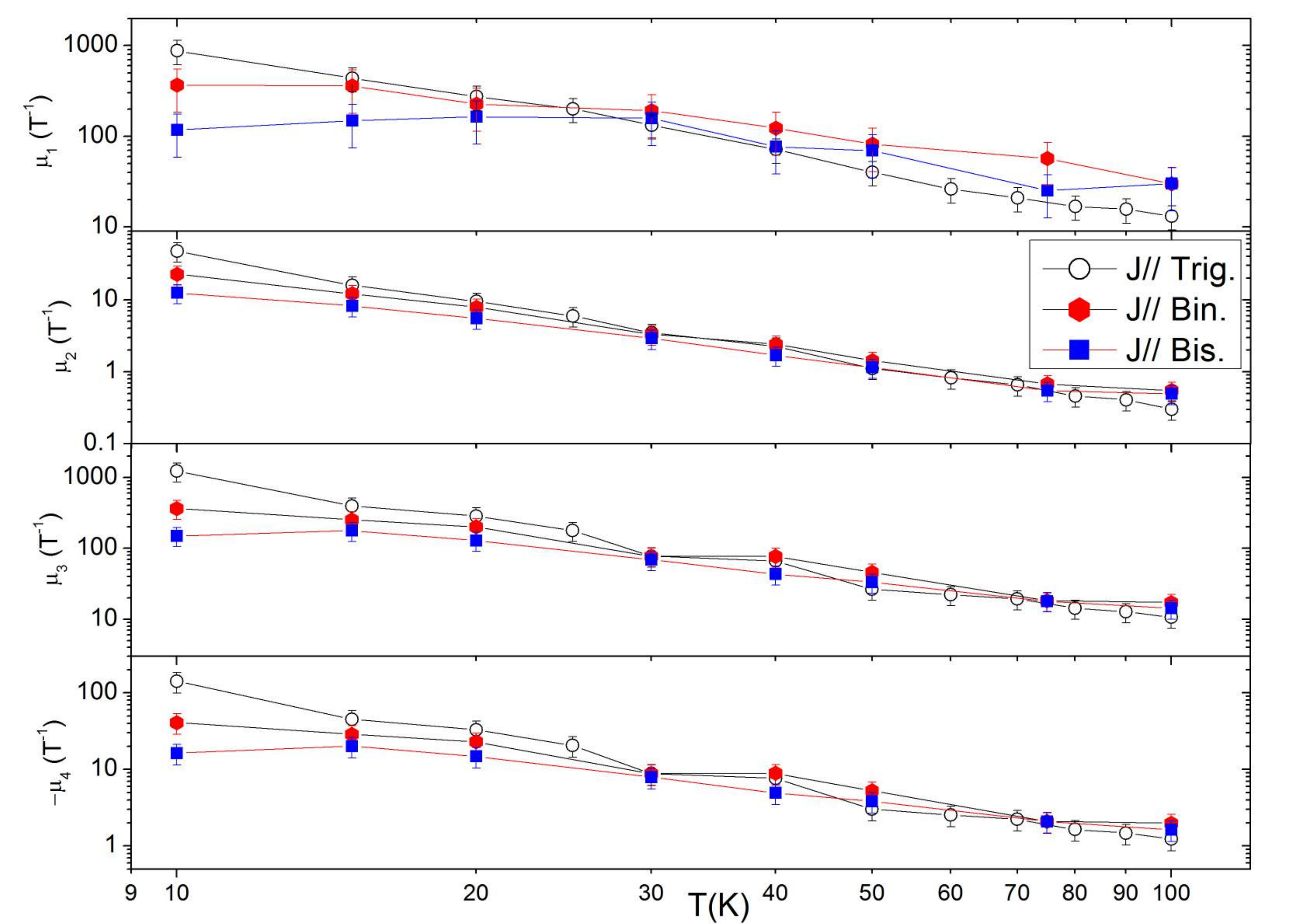}
\caption{Temperature dependence of the components of the electron mobility tensor, $\mu_{i}$, obtained by fitting our angle-dependent magnetoresistance data at $B=0.5$ T for the three planes of rotation. The discrepancy increases with decreasing temperature.}
\end{figure}

One aspect of our data remains currently beyond our understanding. In the first configuration, in which the current is applied along the trigonal axis and  the field is rotating in the (binary, bisectrix) plane, the contribution of the hole pocket is expected to be independent of the orientation of magnetic field. Indeed, for this configuration, the hole conductivity becomes simply:

\begin{equation}
\sigma_{33}^{h} = \frac{3ne\nu_{3}}{1+\nu_{1}\nu_{3}B^{2}}
\end{equation}

Therefore, the fit to the experiment at a fixed magnetic field B yields only a single parameter proportional to the right hand side of this equation and not an independent estimation of $\nu_{1}$ and $\nu_{3}$. The mystery is that the $\nu_{1}$ and the $\nu_{1}\nu_{3}$ found from the data obtained in this configuration are five times larger than what was found for the two other configurations. This discrepancy is well beyond our uncertainty margin. It remains the main failure of the semi-classical model employed here and points to an additional transport process not taken into account in the picture employed here.

\section{Emergence of a valley-polarized state}
\begin{figure}
\includegraphics[width=8.5cm]{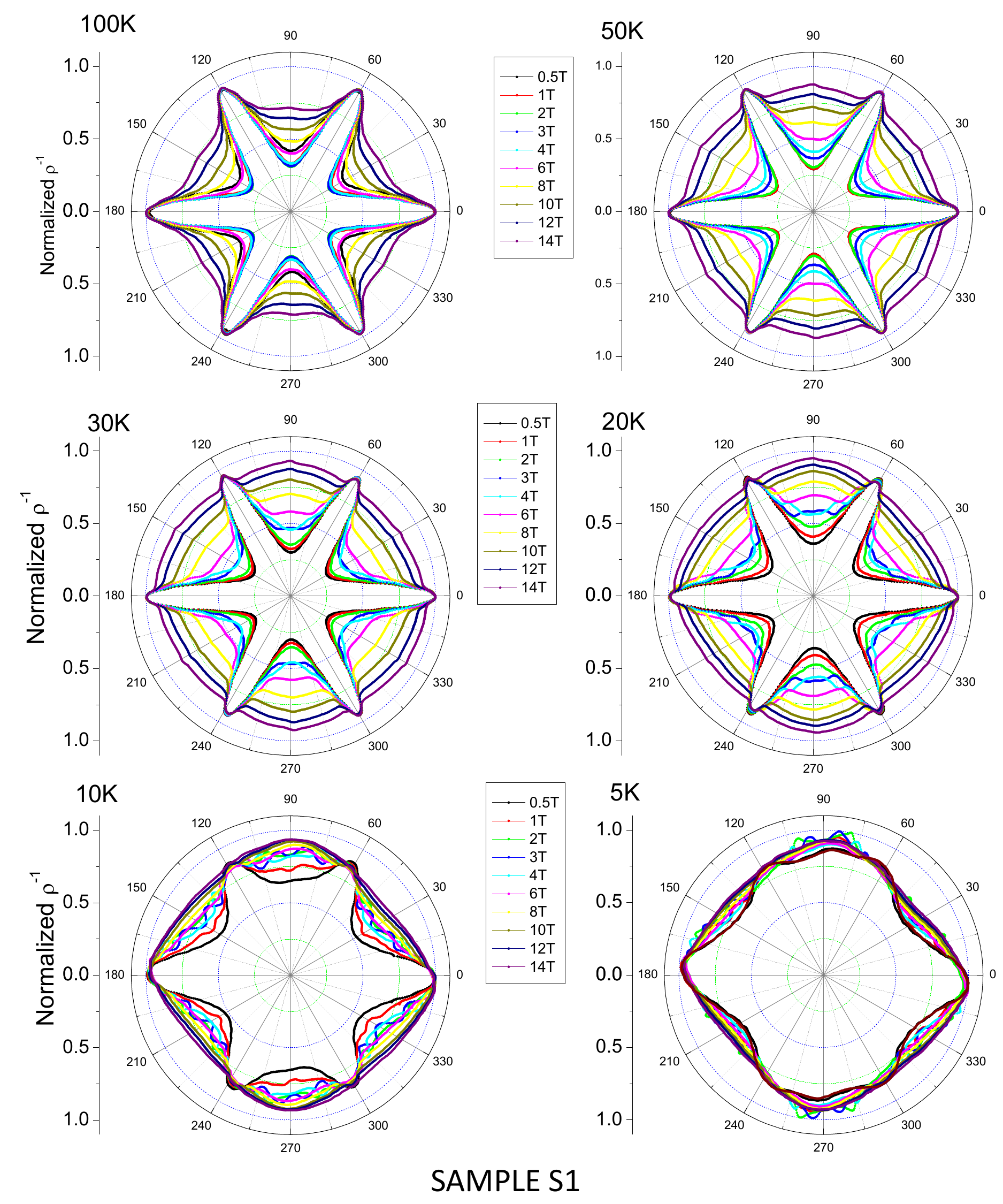}
\caption{Evolution of the angle-dependent magneto-conductivity with decreasing temperature. Each panel shows a polar plot of the inverse of magnetoresistance normalized to its maximum value at a given magnetic field. At high-temperature, the threefold symmetry of the underlying lattice is preserved. As the temperature decreases, this symmetry is lost above a threshold magnetic field, which decreases in amplitude with cooling.}
\end{figure}

Let us now turn our attention to the loss of valley symmetry, a feature reported in two previous experimental studies\cite{zhu2012b,kuechler2014}. In order to explore this feature, we have performed extensive measurements with numerous samples performed in two different laboratories (Paris and Seoul) to document the emergence of spontaneous valley polarization.

When an electric current is applied along the trigonal axis, the three electron valleys remain degenerate. Of course, this degeneracy is lifted by a magnetic field applied along an arbitrary orientation in the (binary, bisectrix) plane.  However, the  C$\overline{3}$ symmetry of the system implies that a $2\pi/3$ rotation of the magnetic field around the trigonal axis should not affect the physical properties, since the degeneracy remains lifted in exactly the same way. Therefore, a magnetic field rotating in the (binary, bisectrix) plane, should generate an angle-dependent magnetoresistance keeping a six-fold symmetry in a polar plot, resulting from the combination of the C3 symmetry and the inversion symmetry. This is indeed the case of our data at high temperature and/or low magnetic field. At low temperature and high magnetic field, on the other hand, this symmetry was found to be lost in all the samples studied.

\begin{figure}
\includegraphics[width=8.5cm]{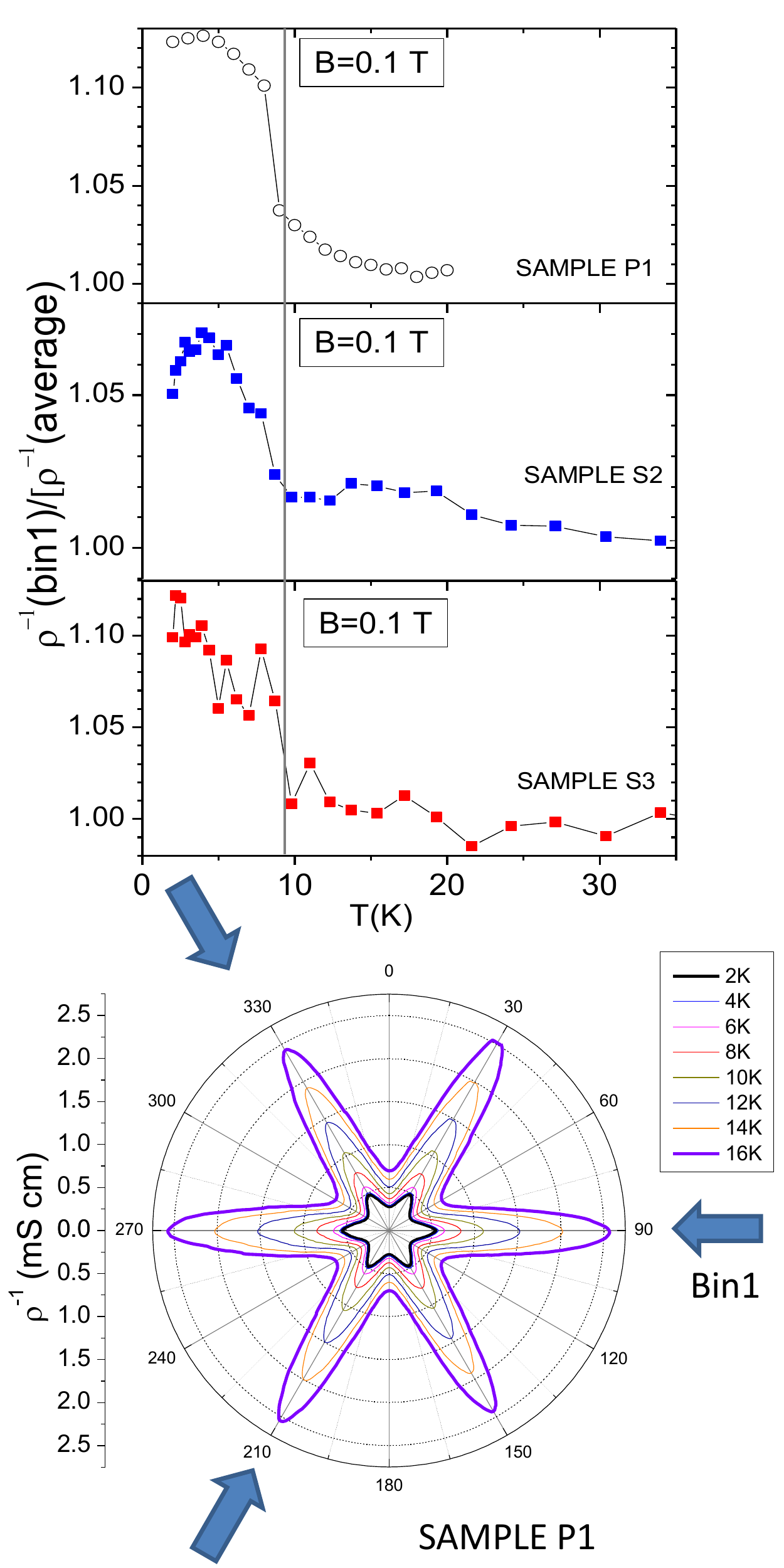}
\caption{Top: The ratio of $\rho^{-1}_\mathrm{{bin1}}$ to $\frac{1}{3}(\rho^{-1}_{\mathrm{bin1}}+\rho^{-1}_{\mathrm{bin2}}+\rho^{-1}_{\mathrm{bin3}})$ as function of temperature in three different single crystals of bismuth at $B=0.1$ T. In all three cases, a sudden jump occurs at $T\sim $9 K. Bottom: polar plots of $\rho^{-1}$ for sample P1.}
\end{figure}

Fig. 9 presents a detailed study on a bismuth single crystal with a square cross-section. Each panel of the figure shows polar plots of $\rho^{-1}$ at a given temperature for different magnetic fields. In the following, $\rho^{-1}$ would be called ``conductivity'' keeping in mind that, because of the contribution of the off-diagonal elements  $\sigma \neq \rho^{-1}$. At each temperature, the data is normalized to the maximum value of $\rho^{-1}$ to allow an easy examination of the evolution as the temperature decreases and the magnetic field increases. One can see that the sixfold symmetry clearly present at 100 K for all magnetic fields is lost at 5K for all magnetic fields. In the intermediate temperature range, the sixfold symmetry is present at low magnetic field but is lost at higher fields.

The loss of threefold symmetry with cooling, originally reported in ref.\cite{zhu2012b}, has been confirmed in more than ten different crystals of bismuth with different shapes and cut from three different mother single crystals and using different experimental set-ups in Paris and in Seoul. The low-temperature asymmetries found in various samples were different from one another. In particular, the departure from threefold symmetry is more striking in samples with a square cross-section than in samples with a cylindrical or a triangular cross-section. On the other hand, the temperature and magnetic field thresholds for the loss of symmetry were roughly similar.

\begin{figure}
\includegraphics[width=7.5cm]{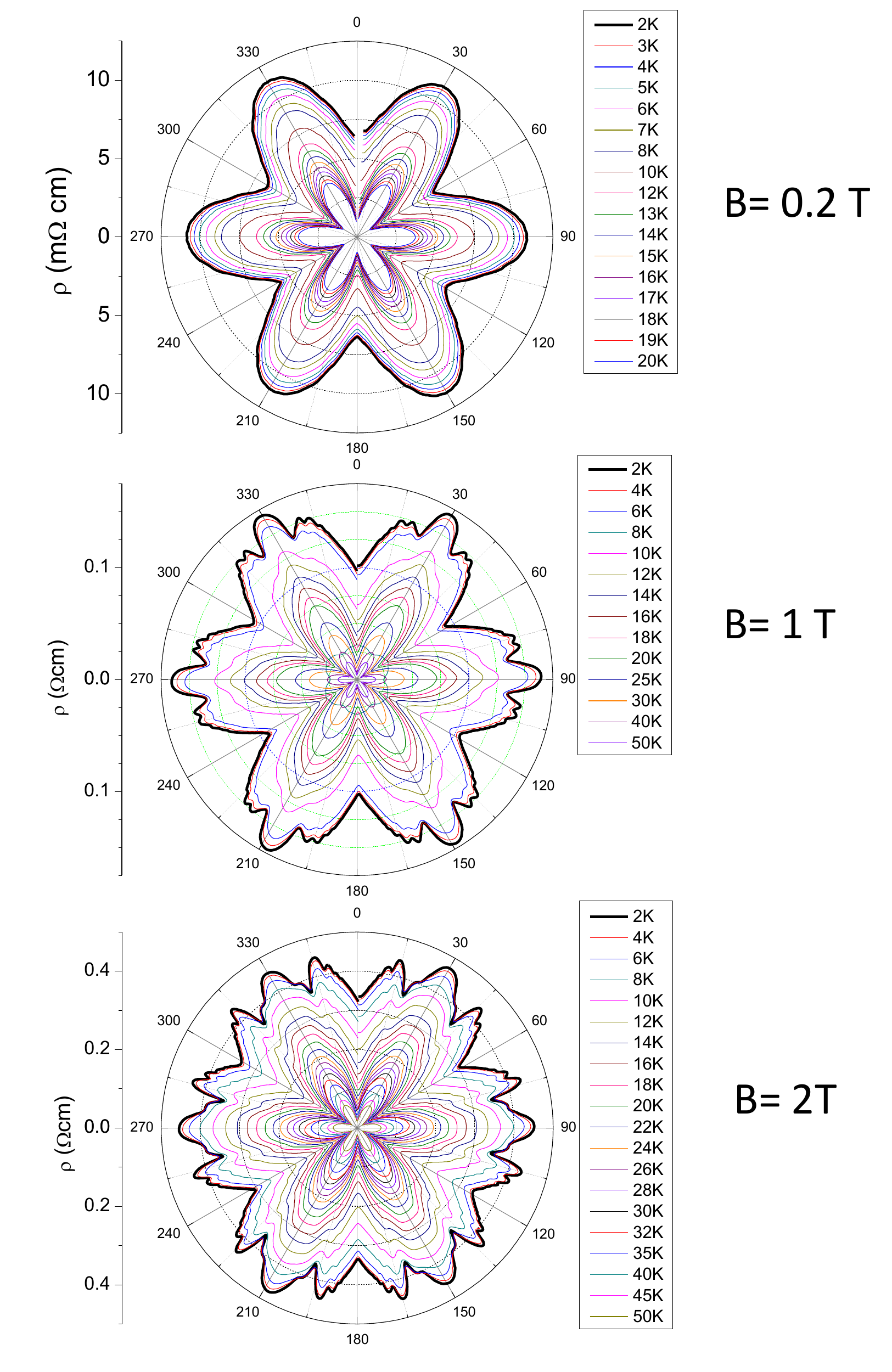}
\includegraphics[width=7.5cm]{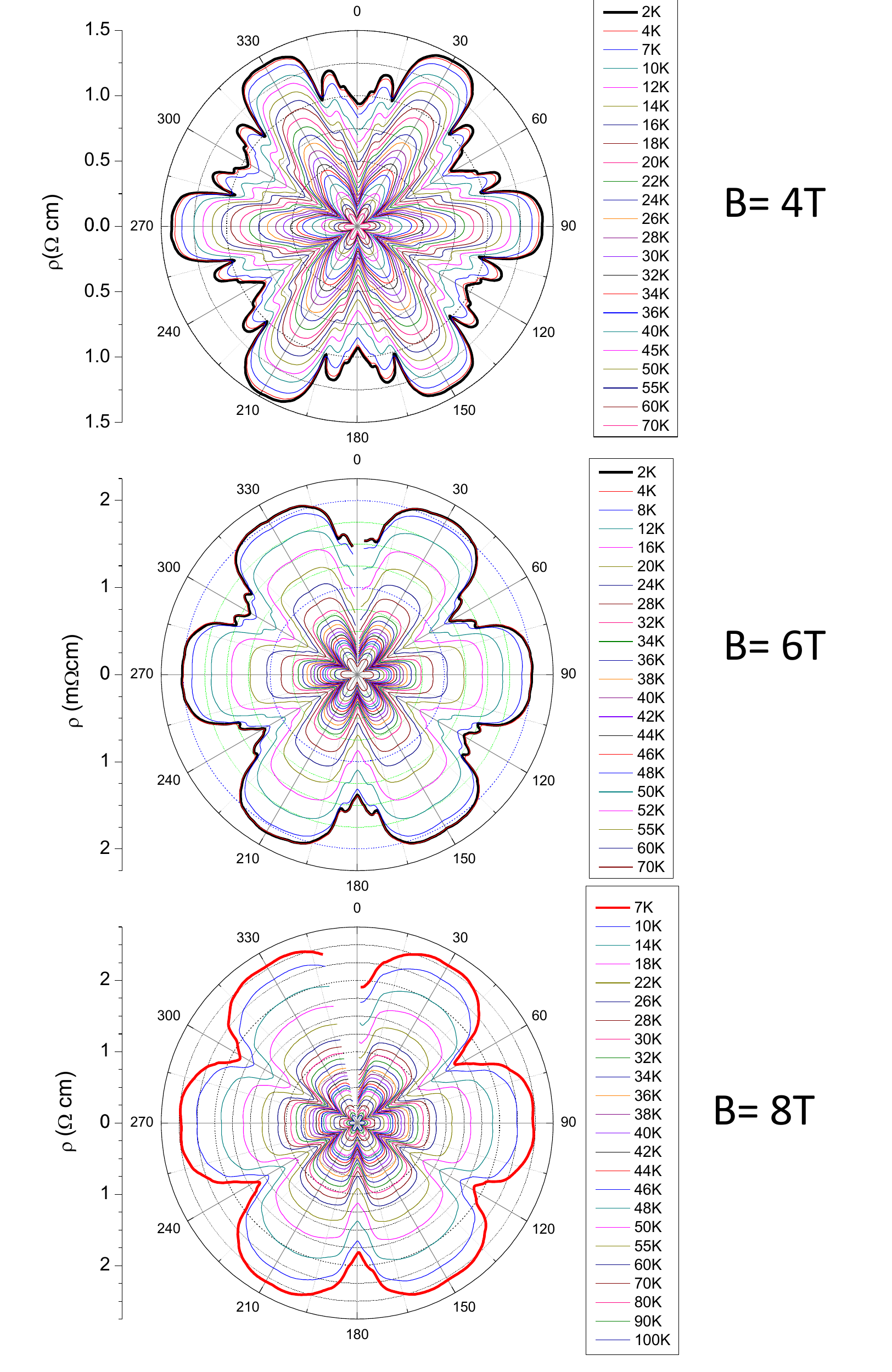}
\caption{Polar plots of resistivity for sample P1 at different temperatures. Each panel shows a different magnetic field. An additional sub-structure arises by occupation and evacuation of Landau levels with the rotating magnetic field. The angular pattern becomes simple again at 8 T as no more Landau level is occupied or evacuated during the rotation. }
\end{figure}

Various possible experimental artifacts as the source of the loss of threefold symmetry can be ruled out. Two-axis rotation experiments performed have found a similar loss of threefold symmetry\cite{zhu2012b}. Therefore, it cannot be a consequence of uncontrolled misalignment between the magnetic field, the crystal axes and applied electric current. The fact that the frequency of quantum oscillations remains identical for different orientations of magnetic field\cite{kuechler2014}(see also below) rules out any possible role of internal uncontrolled strain, since experiments under strain have shown that presence of strain would have caused a detectable variation among the three electron pockets\cite{brandt1980}. Our bismuth crystals are twinned and  therefore in addition to the dominant domain, there are three minority domains. The presence of these minority domains generates  a secondary set of Landau peaks visible by the Nernst measurements\cite{zhu2012a}. But, we cannot think of anyway their presence can distort the threefold symmetry of the dominant domain, the only source of magnetoresistance in our study.

\begin{figure}
\includegraphics[width=10cm]{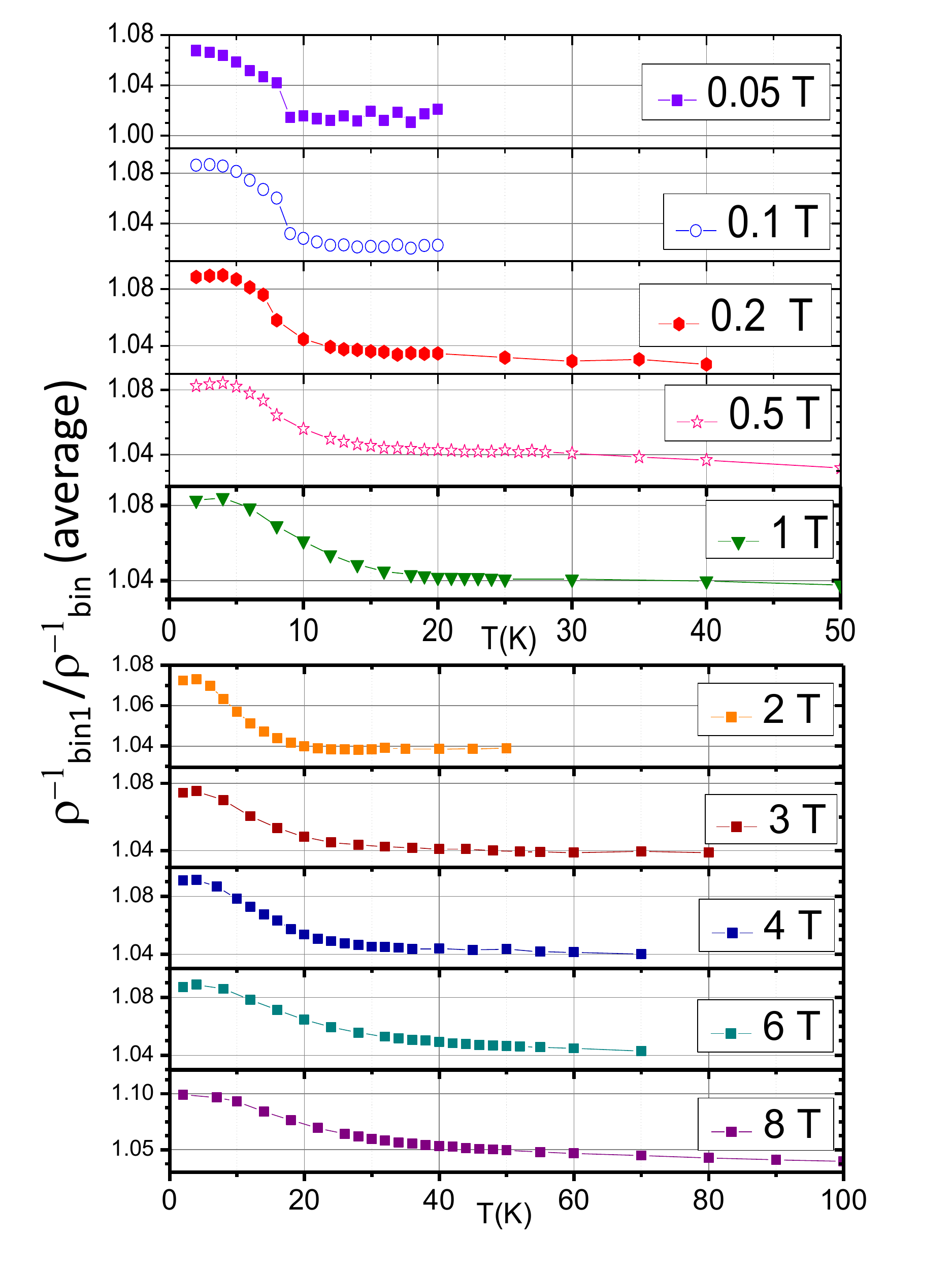}
\caption{The evolution of the jump observed in the ratio of $\rho^{-1}_{bin1}$ to $\frac{1}{3}(\rho^{-1}_{bin1}+\rho^{-1}_{bin2}+\rho^{-1}_{bin3})$ (see Fig. 10) with increasing magnetic field in sample P1 using the data plotted in Fig. 11 . As the magnetic field increases, the transition becomes wider and shifts to higher temperatures. }
\end{figure}

By following the evolution of this loss of symmetry, we succeeded to identify a visible signature of a phase transition associated with this loss of symmetry at low magnetic fields. It is in the low-field window that an abrupt change becomes visible. As seen in Fig. 10, at $B = 0.1$ T, the magnitude of magnetoresistance for magnetic field aligned along one bisectrix axis suddenly deviates from its average value at a temperature. Because of the residual experimental misalignment, the amplitude of resistivity for magnetic field oriented along each of the three bisectrix axes is slightly different. However, as seen in the figure, this departure from unity is almost constant at higher temperature. Suddenly, below a field-dependent threshold temperature, an abrupt deviation on top of the experimental misalignment can be clearly detected. As seen in the figure, in two other samples studied at $B = 0.1$ T, a similar abrupt jump occurring at an almost identical temperature was found.  This implies that the boundary marking the loss of threefold symmetry is an intrinsic property of bismuth.

In two samples P1, a cylinder with a circular cross section, and P2, a cuboid with a cleaved square cross section, we made extensive measurements up to 8 T in order to follow the evolution of this jump. The data for sample P1 is shown in Fig. 11.  As the magnetic field rotates, Landau levels are filled and evacuated and this adds additional substructure to the background. This substructure evolves as the magnetic field increases and finally vanishes at 8 T since no more filling or evacuation occurs. Nevertheless, the jump associated with the loss of threefold symmetry can be extracted by the same procedure. The result is shown in Fig. 12. There are two distinct visible effects of the magnetic field on the phase transition. With increasing magnetic field, the transition shifts to higher temperatures and becomes wider. In a field as high as 8 T, the loss of symmetry occurs over a temperature range as wide as 20 K. On the other hand, the magnitude of the maximum deviation from unity does not change with magnetic field and remains a few percent in the whole range of investigation.

Using this data, we can construct a phase diagram drawing the boundaries of the state in which the symmetry of the three valleys is lost. This phase diagram is shown in Fig. 13. Each symbol represents the position of the intersection between the low-temperature and high-temperature behaviors detected in Fig. 12. The C3 symmetry is preserved in the low-field-high-temperature region and lost in the high-field-low-temperature region. A very similar phase diagram is found for sample P2 which is a cuboid with a square cross section. As seen in the figure, while the cross section does have a signature in the ultimate pattern of angular magnetoresistance in the low-temperature-high-field state, it does not modify the boundary in a significant way. This provides further evidence that this transition is intrinsic to bulk bismuth. We did not find any evidence for hysteresis in these measurements.

\begin{figure}
\includegraphics[width=9cm]{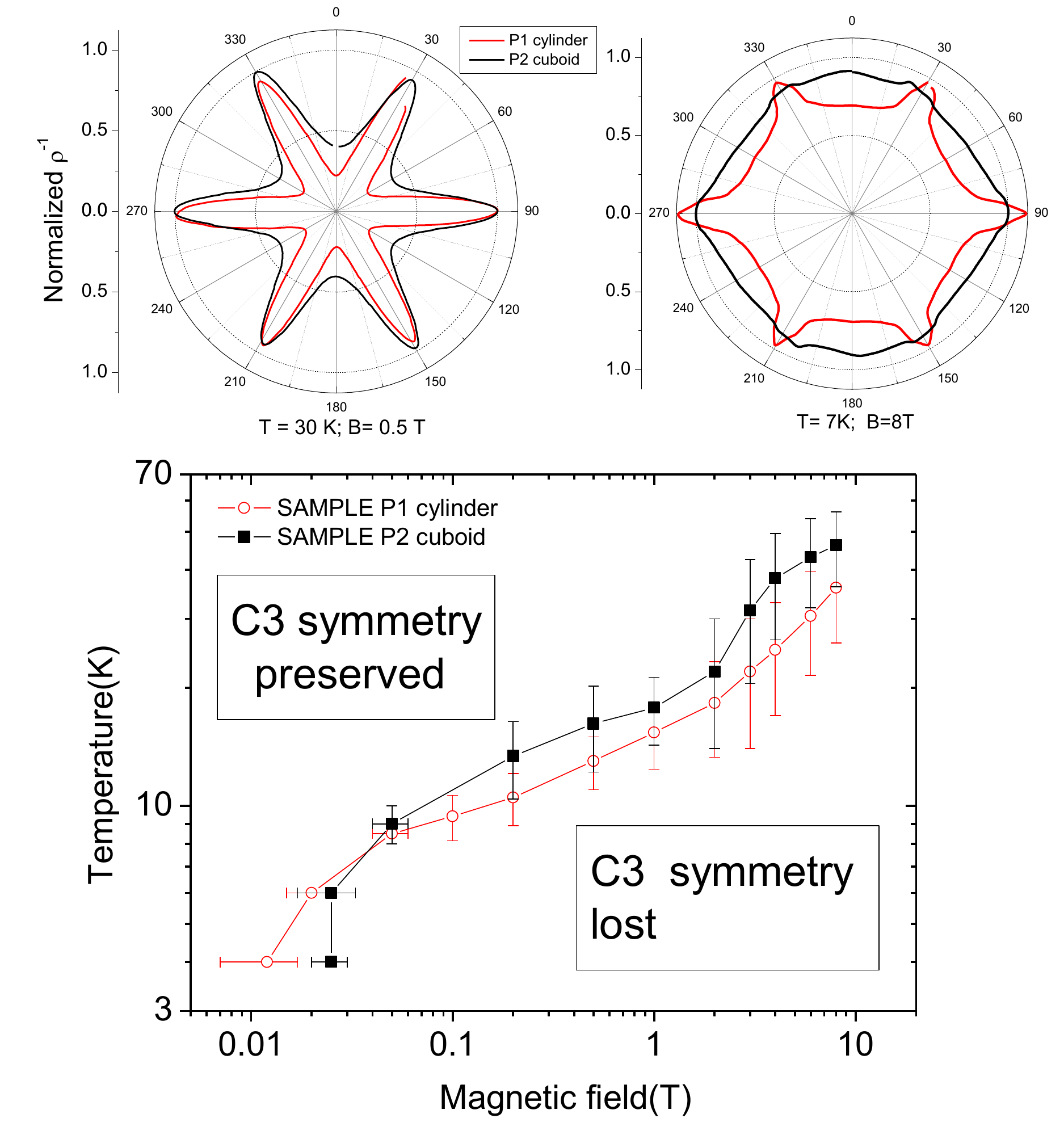}
\caption{Top: Comparison of the angular variation of normalized conductance in two samples with different cross sections. In the high-temperature-low-field state, they both display the threefold symmetry, but in the low-temperature-high-field state they loose it with different patterns emerging in the valley-polarized state. Bottom: Variation of the critical temperature for the loss of threefold symmetry with magnetic field in the two samples. Vertical (horizontal) error bars indicate the width of transition as the temperature(magnetic field) is swept at a fixed magnetic field (temperature). The boundary between the ordered (less symmetric) and disordered (more symmetric) states in the (field, temperature) plane is similar.}
\end{figure}

We also studied the quantum oscillations of resistivity (the Shubnikov- de Haas effect) for different orientations of magnetic field along the three binary and the three bisectrix axes . We found that while the frequency of oscillations is identical for the three equivalent orientations, there is a detectable difference in the amplitude of oscillations. This result is similar to what was reported by K\"{u}chler \emph{et al.} in the case of the quantum oscillations of magnetostriction\cite{kuechler2014} and the quantum oscillations of the Nernst coefficient\cite{yang2010}. While there is no visible difference between the angle-resolved Landau spectrum of the three electron valleys, the intensity of the signal generated by emptying a Landau level is different among valleys. As seen in Fig. 14, the amplitude of oscillations obtained after subtracting a smooth background are different for three nominally equivalent orientations of magnetic field. However, the oscillations have identical frequencies. Moreover, there is no visible difference in the oscillations of the second-derivative of resistivity, which wipes out any field-linear magnetoresistance. Therefore, it is tempting to conclude that the difference in amplitude of quantum oscillations for the three valleys is roughly linear in magnetic field, like the monotonous background.

To sum up, this study confirms the emergence of valley polarization in low temperature and high-magnetic field reported previously\cite{zhu2012b,kuechler2014}. In addition and for the first time, an abrupt change reminiscent of a phase transition can be clearly seen in the data and the boundary of the valley-polarized state (for a magnetic field perpendicular to the trigonal axis) can be established by following the evolution of this anomaly with field and temperature. Thus, there is now a body of experimental results pointing to valley polarization in bismuth, a phenomenon not still understood. While several theoretical ideas have emerged, none of them offers a satisfactory explanation of the whole spectrum of the observations.

One appealing line of thought is to follow the idea of valley nematicity proposed for quantum Hall systems with multiple eccentric valleys\cite{abanin2010}. Abanin and co-workers argued that in quantum Hall systems, i.e. two-dimensional systems subject to quantizing magnetic fields, if the valleys present an anisotropic effective mass, unequal occupation of valleys would save exchange energy among electrons. As a consequence, a ground state in which one of the valleys is preferably occupied by electrons, dubbed a valley-nematic state, can be induced by magnetic field. In this scenario, the Coulomb interaction and the mass anisotropy in each valley play an important role. The scenario is conceived mainly with a two-dimensional system with C4 symmetry such as AlAs heterostructures\cite{shkolnikov2005}. Bulk bismuth has a C3 symmetry and is three-dimensional. On the other hand, it has a mass anisotropy forty times larger than AlAs, making it a potential candidate for this scenario.

Experiment, however, indicates that the three electron-like pockets do not differ in their size. The identical frequency of quantum oscillations implies identical valley occupation. This is not conform to the expectations along this line of thought. The three valleys may differ in the density of states near the chemical potential. This is what governs the magnitude of quantum oscillations as well as the number of electrons participating in charge transport. But, how can the three valleys remain identical in size, but be different in their  density-of-states near the chemical potential? There is no satisfactory answer to this question. However, K\"{u}chler \emph{et al.} \cite{kuechler2014} recall the case of disordered semiconductors, in which the combination of Coulomb interaction and disorder open a gap in the immediate vicinity of the chemical potential\cite{efros1975}. It is still unclear how one can use this idea in the case of a multi-valley metal such as bismuth.

Another possibility is a field-induced lattice distortion driven by electron-phonon coupling. This idea, proposed by Mikitik and Sharlai\cite{mikitik2014}, is reminiscent of Jahn-Teller distortion in insulators\cite{jahn1937}. According to the Jahn-Teller theorem, a number of exceptions notwithstanding, a geometrical configuration of atoms becomes unstable in presence of electron degeneracy\cite{knox1964}. Lowering symmetry allows the molecule to get rid of the costly electronic degeneracy. In the case of bismuth, the degeneracy of the three valleys has an energy cost, which may be reduced thanks to lattice distortion. There is currently no  evidence for such a field-induced lattice distortion. There has also been no attempt to pin it down if it exists.

Can a phase transition with a critical temperature of 9 K be generated by a magnetic field as small as 0.1 T? To address this question, one should recall that the cyclotron energy is inversely proportional to the mass of electrons. While, for free electrons, the cyclotron energy at 0.1 T is only 0.13 K,  in bismuth, where electrons are one thousand times lighter along the bisectrix, the cyclotron energy becomes orders of magnitude larger at the same magnetic field and a small magnetic field can cause a relatively robust instability.

\begin{figure}
\includegraphics[width=9cm]{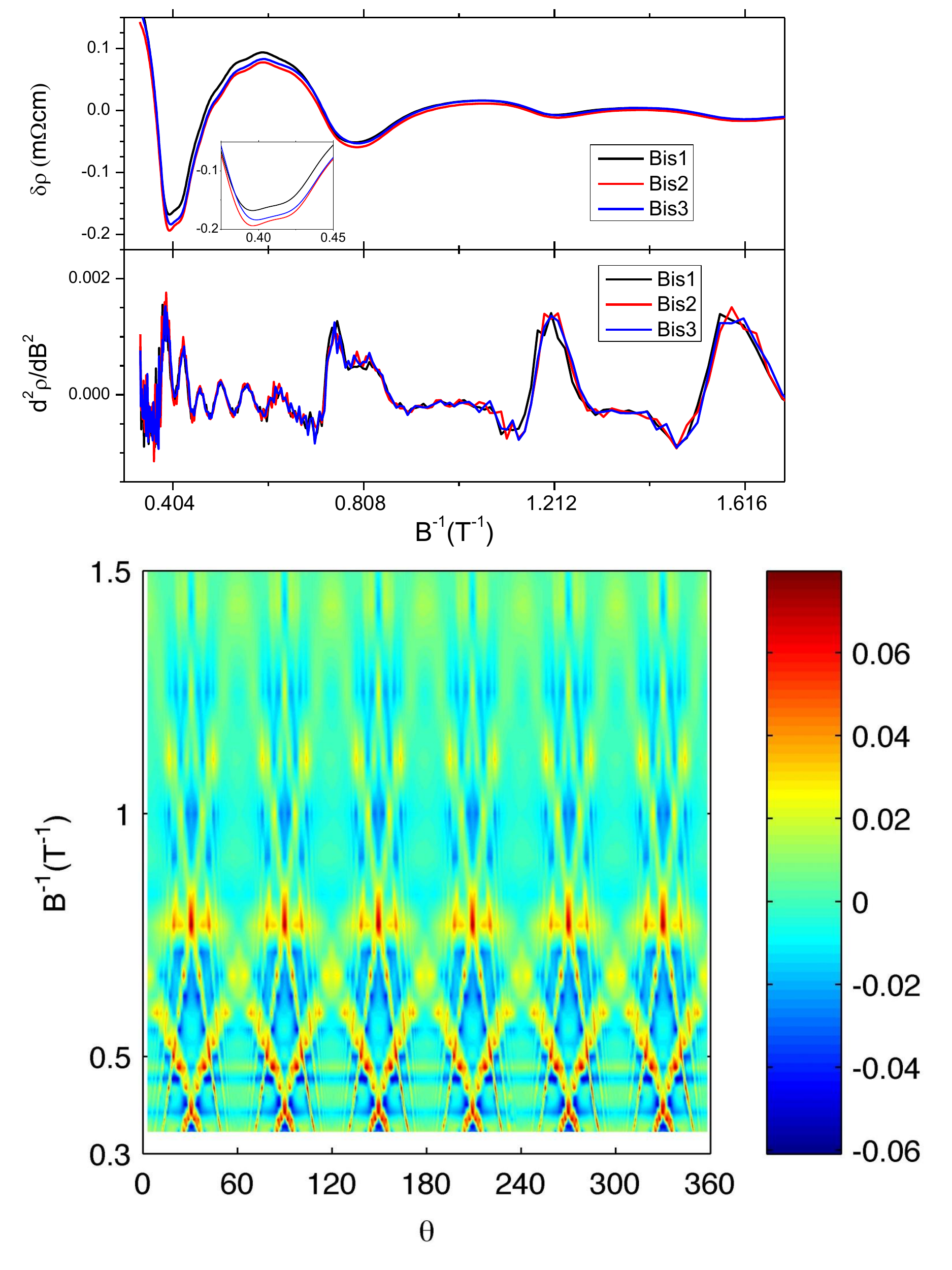}
\caption{Top: Comparison of quantum oscillations of magnetoresistance (the Shubnikov-de Haas effect) for magnetic field along three equivalent bisectrix axes. The amplitude of the oscillating signal, obtained after subtracting a smooth background for each angle is different along the three orientations. The difference becomes undetectable in the second derivative of magnetoresistance along the three different orientations. Bottom: The angle-resolved Landau spectrum revealed in a color plot of d$^{2}\rho$/dB$^{2}$ keeps the threefold symmetry.}
\end{figure}
\section{Concluding remarks}

Decades ago, Wolff wrote: ``The metal  bismuth  is  one  of  the  most  peculiar  and intriguing  of  solids''\cite{wolff1964}. This peculiarity shows itself in numerous physical properties ranging from its large diamagnetism to its remarkable thermoelectric figure of merit. In the case of our study, three basic facts are relevant: i) Carriers in bismuth are extremely mobile; ii) This mobility is extremely anisotropic; iii) The anisotropy is inverted for electron-like and hole-like carriers. There are other solids displaying these features, but the magnitude of mobility as well as the intricate complexity of its anisotropy puts bismuth in a league of its own.

The first part of this study documents the evolution of angle-dependent magnetoresistance with temperature and magnetic field and finds that the complex angular variation is within the reach of a semi-classical treatment if the mobility is taken as a tensor. This paves the way for a quantitative understanding of magnetoresistance in bismuth at arbitrarily large fields.  Upon the application of a magnetic field as large as 65 T along the trigonal axis, bismuth remains metallic with an electric resistance increasing by six orders of magnitude\cite{fauque2009}. This million-fold increase is even larger than what has been recently reported in WTe$_{2}$\cite{ali2014} in a comparable field range. However, in contrast to the latter and what is expected for a simple compensated metal\cite{pippard1989},  the  magnetoresistance does not present a quadratic field-dependence and cannot be described by a simple power law over an extended field window.  One identified reason for this deviation is the field-induced change in carrier density at high magnetic fields. When the magnetic field exceeds 0.5 T, the degeneracy of Landau bands leads to a modification of the carrier number, while preserving the charge conservation. This feature, which has visible signatures in the high-field Landau spectrum\cite{zhu2011}, should be taken into account in any realistic description of high-field magnetoresistance. Quantifying the components of mobility of electrons and holes in the low-field limit is a beginning. Understanding  magnetoresistance at arbitrarily large magnetic field cannot be achieved by a toy model of a compensated semi-metal with a scalar mobility. Future studies will tell if one can describe the field dependence of magnetoresistance with realistic assumptions on the amplitude of the mobility components and their field dependence.

The second part of this study documents the emergence of a valley-polarized state at low temperature and high magnetic field. Currently, this state, in which the electron fluid loses the threefold rotational symmetry of the underlying lattice, is poorly understood.  Nematicity, a subject of intense attention in strongly-correlated electron systems, is associated with the loss of rotational symmetry by the electron fluid\cite{fradkin2010}. It is tempting, and maybe even legitimate, to qualify the valley-polarized state found in bismuth as valley-nematic. However, one shall not forget that the picture drawn by experimental results does not easily fit in the most straightforward description of valley nematicity. In the ordered state, the three electron valleys remain identical in size but differ in their density of states at the Fermi level. There is no satisfactory understanding of how this can be achieved.

To sum up, elemental bismuth hosts extremely mobile and very anisotropic carriers, which scatter off each other in a complex, yet increasingly documented, manner. This looks like an appealing platform to explore the possible existence of electronic equivalents of liquid crystals.

\section{Acknowledgements}

This work was supported in France  by the QUANTHERM and SUPERFIELD projects funded by \emph{Agence Nationale de la Recherche} and a grant attributed by the \emph{Ile de France} regional council. Support by France-Japan Sakura and France-Korea STAR collaboration programs is also acknowledges. W. K. is supported by the NRF grants funded by the Korea Government (MSIP) (No. 2015-001948 and No. 2010-00453).


\begin{thebibliography}{}
\bibitem{kapitza1928} P. Kapitza, The Study of the Specific Resistance of  Bismuth Crystals and Its Change in Strong Magnetic Fields and Some Allied Problems, Proc. R. Soc. A \textbf{119}, 358 (1928)
\bibitem{edelman1976} V. S. Edelman, Electrons in bismuth, Adv. Phys. \textbf{25}, 555 (1976)
\bibitem{fuseya2014} Y. Fuseya, M. Ogata and H. Fukuyama, Transport Properties and Diamagnetism of Dirac Electrons in Bismuth, J. Phys. Soc. Jpn. \textbf{84}, 012001 (2015)
\bibitem{behnia2007} K. Behnia, L. Balicas, Y. Kopelevich, Signatures of electron fractionalization in ultraquantum bismuth, Science \textbf{317}, 1729 (2007)
\bibitem{luli2008} L. Li, J. G. Checkelsky, Y. S. Hor, C. Uher, A. F. Hebard, R. J. Cava and N. P. Ong, Phase transitions of Dirac electrons in bismuth, Science \textbf{321}, 547 (2008)
\bibitem{fauque2009} B. Fauqu\'e, B. Vignolle, C. Proust,  J.-P. Issi and K. Behnia, Electronic instability in bismuth far beyond the quantum limit, New J. Phys. \textbf{11}, 113012 (2009)
\bibitem{fauque2009b} B. Fauqu\'e, H. Yang, I. Sheikin, L. Balicas, J.-P. Issi, and K. Behnia, Hall plateaus at magic angles in bismuth beyond the quantum limit, Phys. Rev. B \textbf{79}, 245124 (2009)
    \bibitem{yang2010} H. Yang, B. Fauqu\'e, L. Malone, A. B. Antunes, Z. Zhu, C. Uher and K. Behnia, Phase diagram of bismuth in the extreme quantum limit, Nat. Commun. \textbf{1}, 47 (2010)
\bibitem{zhu2011} Z. Zhu,  B. Fauqu\'e, Y. Fuseya and K. Behnia, Angle-resolved Landau spectrum of electrons and holes in bismuth, Phys. Rev. B\textbf{ 84}, 115137 (2011)
\bibitem{zhu2012a} Z. Zhu, B. Fauqu\'e, L. Malone, A. B. Antunes, Y. Fuseya and K. Behnia, Landau spectrum and twin boundaries of bismuth in the extreme quantum limit, PNAS \textbf{109}, 14813 (2012)
\bibitem{alicea2009} J. Alicea and L. Balents, Bismuth in strong magnetic fields: Unconventional Zeeman coupling and correlation effects, Phys. Rev. B \textbf{79}, 241101(R)(2009)
\bibitem{sharlai2009} Yu. V. Sharlai and G. P. Mikitik,  Origin of the peaks in the Nernst coefficient of bismuth in strong magnetic fields, Phys. Rev. B \textbf{79}, 081102(R) (2009)
\bibitem{zhu2012b} Z. Zhu, A. Callaudin, B. Fauqu\'e, W. Kang and K. Behnia, Field-induced polarization of Dirac valleys in bismuth,  Nature Physics \textbf{8}, 89 (2012)
\bibitem{kuechler2014} R. K\"{u}chler, L. Steinke, R. Daou,  M. Brando, K. Behnia and F. Steglich, Thermodynamic evidence for valley-dependent density of states in bulk bismuth, Nature Mater. \textbf{13}, 461 (2014)
\bibitem{liang2014} T. Liang, Q. Gibson, M. N. Ali, M. Liu, R. J. Cava, N. P. Ong, Ultrahigh mobility and giant magnetoresistance in Cd$_3$As$_2$: protection from backscattering in a Dirac semimetal, Nature Mater. \textbf{14}, 280 (2015)
\bibitem{ali2014} M. N. Ali, J. Xiong, S. Flynn, J. Tao, Q. D. Gibson, L. M. Schoop, T. Liang, N. Haldolaarachchige, M. Hirschberger, N. P. Ong and R. J. Cava, Large, non-saturating magnetoresistance in WTe$_{2}$, Nature \textbf{514}, 205 (2014)
\bibitem{kopelevich2003} Y. Kopelevich, J. H. S. Torres, R. R. da Silva, F. Mrowka, H. Kempa, and P. Esquinazi,  Reentrant Metallic Behavior of Graphite in the Quantum Limit, Phys. Rev. Lett. \textbf{90}, 156402 (2003)
\bibitem{du2005}  X.  Du, S. -W. Tsai, D. L. Maslov and A. F. Hebard, Metal-insulator-like behavior in semimetallic bismuth and graphite. Phys. Rev. Lett. \textbf{94}, 166601 (2005)
\bibitem{bergemann2000} C. Bergemann, S. R. Julian, A. P. Mackenzie, S. Nishizaki, and Y. Maeno,  Detailed Topography of the Fermi Surface of Sr$_{2}$RuO$_{4}$, Phys. Rev. Lett. \textbf{84}, 2662 (2000)
\bibitem{hussey2003} N. E. Hussey, M. Abdel-Jawad, A. Carrington, A. P. Mackenzie and  L. Balicas, A coherent three-dimensional Fermi surface in a high-transition-temperature superconductor,  Nature \textbf{425}, 814 (2003)
\bibitem{kang2007} W. Kang, T. Osada, Y. J. Jo, and Haeyong Kang,  Interlayer Magnetoresistance of Quasi-One-Dimensional Layered Organic Conductors
Phys. Rev. Lett. \textbf{99}, 017002(2007)
\bibitem{abanin2010} D. A. Abanin, S. A. Parameswaran, S. A. Kivelson, and S. L. Sondhi, Nematic Valley Ordering in Quantum Hall Systems, Phys. Rev. B\textbf{ 82}, 035428 (2010)
\bibitem{mikitik2014} G. P. Mikitik and Yu. V. Sharlai, Spontaneous symmetry breaking of magnetostriction in metals with multi-valley band structure, Phys. Rev. B \textbf{91}, 075111 (2015)
\bibitem{liu1995} Y. Liu and R. E. Allen, Electronic structure of the semimetals Bi and Sb, Phys. Rev. B \textbf{52}, 1566 (1995)
\bibitem{mase1962} S. Mase, S. von Molnar, and A. W. Lawson,  Galvanomagnetic tensor of bismuth at 20.4K, Phys. Rev. \textbf{127}, 1030 (1962)
\bibitem{rycerz2007} A. Rycerz, J. Tworzydlo and C. W. J. Beenakker, Valley filter and valley valve in graphene, Nature Phys. \textbf{3}, 172  (2007)
\bibitem{behnia2012} K. Behnia, Condensed-matter physics: Polarized light boosts valleytronics, Nature Nanotech. \textbf{7}, 488 (2012)
\bibitem{jo2014} Y. J. Jo, Joonbum Park, G. Lee, Man Jin Eom, E. S. Choi, Ji Hoon Shim, W. Kang, and Jun Sung Kim, Valley-Polarized Interlayer Conduction of Anisotropic Dirac Fermions in SrMnBi$_{2}$, Phys. Rev. Lett. \textbf{113}, 156602 (2014)
\bibitem{abeles1956} B. Abeles and S. Meiboom, Galvanomagnetic effects in bismuth, Phys. Rev. \textbf{101}, 544 (1956)
\bibitem{aubrey1971} J. E. Aubrey, Magnetoconductivity tensor for semimetals, J. Phys. F \textbf{1}, 493 (1971)
\bibitem{Ziman1960} J. M. Ziman, Electrons and Phonons, Clarendon Press, Oxford (1960)
\bibitem{smith1967} A. C. Smith, J. F. Janak and R. B. Adler, Electronic Conduction in Solids, McGraw-Hill, New York (1967)
\bibitem{dresselhaus1971} M. S. Dresselhaus. Electronic properties of the group V semimetals, J. Phys. Chem. Solids \textbf{32} ( Suppl. 1),  3 (1971)
\bibitem{akgoz1975} Y. C. Akg\"{o}z  and G. A. Saunders, Space-time symmetry restrictions on the form of transport tensors. I. Galvanomagnetic effects, J. Phys. C \textbf{8}, 1387 (1975)
\bibitem{sumengen1974} Z. S\"{u}mengen, N. T\"{u}retken, and G. A. Saunders, The angular dependence of the magnetoresistivity of bismuth, J. Phys. C \textbf{7}, 2204 (1974)
\bibitem{hartman1969} R. Hartman, Temperature dependence of the low-field galvanomagnetic coefficients of bismuth. Phys. Rev. \textbf{181}, 1070 (1969)
\bibitem{michenaud1972} J. -P. Michenaud and J. -P. Issi, Electron and hole transport in bismuth, J. Phys. C \textbf{5}, 3061 (1972)
\bibitem{brown1968} R. D. Brown, R. L. Hartman, and S. H. Koenig, Tilt of the Electron Fermi Surface in Bi, Phys. Rev. \textbf{172}, 598 (1968)
\bibitem{bhargava1967}R. N. Bhargava, de Haas-van Alphen and Galvanometric effects in Bi and Bi-Pb alloys, Phys. Rev. \textbf{156}, 785 (1967)
\bibitem{issi1979} J. P. Issi,  Low temperature transport properties of the group V semimetals, Aust. J. Phys. \textbf{32}, 585 (1979)
\bibitem{baber1937} W. G. Baber, The contribution to the electrical resistance of metals from collisions between electrons, Proc. Royal Soc. A \textbf{158}, 383 (1937)
 \bibitem{kadowaki1986} K. Kadowaki and S. B. Woods, Universal relationship of the resistivity and specific heat in heavy-fermion compounds,  Solid State Commun. \textbf{58},  507 (1986)
 \bibitem{brandt1980} N. B. Brandt, V. A. Kui'bachinskil, N. Ya. Minina, and V. D. Shirokikh, Change of the band structure and electronic phase
transitions in Bi and Bi$_{1-x}$Sb$_{x}$  alloys under uniaxial tension strains, Sov. Phys. JETP \textbf{51}, 562 (1980)
\bibitem{shkolnikov2005} Y. P. Shkolnikov, S. Misra, N. C. Bishop, E. P. De Poortere, and M. Shayegan, Observation of Quantum Hall “Valley Skyrmions”, Phys. Rev. Lett. \textbf{95}, 066809 (2005)
\bibitem{efros1975} A. L. Efros and B. I. Shklovskii, Coulomb gap and low temperature conductivity of disordered systems,  J. Phys. C: Solid State Phys. \textbf{8}, L49  (1975)
\bibitem{jahn1937}H. A. Jahn and E. Teller, Stability of Polyatomic Molecules in Degenerate Electronic States I-Orbital degeneracy, Proc. Royal Soc. \textbf{A161}, 220 (1937)
\bibitem{knox1964}R. S. Knox and A. Gold, Symmetry in the solid state, W. A. Benjamin, New York (1964)
\bibitem{wolff1964} P.  A.  Wolff, Matrix  elements  and  selection  rules  for  the two-band  model  of  bismuth, J.  Phys.  Chem.  Solids  \textbf{25},  1057 (1964)
\bibitem{pippard1989} A. B. Pippard , Magnetoresistance in metals, Cambridge University Press, Cambridge (1989)
\bibitem{fradkin2010}E. Fradkin, S. A. Kivelson, M. J. Lawler, J. P. Eisenstein and A. P. Mackenzie, Nematic Fermi Fluids in Condensed Matter Physics, Annual Review of Condensed Matter Physics \textbf{1},  153  (2010)

\end{thebibliography}
\end{document}